\documentclass[twocolumn,showpacs,preprintnumbers,amsmath,amssymb]{revtex4}
\usepackage{graphicx}
\usepackage{amssymb}
\usepackage[latin1]{inputenc}
\usepackage{dcolumn}
\usepackage{bm}
\usepackage{amsmath}
\usepackage{mathtools}
\usepackage{textcomp}
\usepackage{amsbsy}
\usepackage[dvips]{color}
\usepackage{float}
\usepackage{times}
\DeclareGraphicsExtensions{.png,.eps}
\newcommand{\glzero}{$\Gamma-{L}_0$ }
\newcommand{\glone}{$\Gamma-{L}_1$ }
\newcommand{\gltwo}{$\Gamma-{L}_2$ }
\newcommand{\glthree}{$\Gamma-{L}_3$ }

\begin{document}



\title{Spin Stiffness and Domain Walls in Dirac-Electron Mediated Magnets}

\author{Sahinur Reja$^{1}$, H.A.Fertig$^{1}$, and L. Brey$^{2}$}
\affiliation{
$^1$Department of Physics, Indiana University, Bloomington, IN 47405\\
$^2$ Instituto de Ciencia de Materiales de Madrid, (CSIC),
Cantoblanco, 28049 Madrid, Spain
}

\date{\today}


\begin{abstract}
Spin interactions of magnetic impurities mediated by conduction electrons is one
of the most interesting and potentially useful routes to ferromagnetism in condensed matter.
In recent years such systems have received renewed attention due to the advent of materials
in which Dirac electrons are the mediating particles, with prominent examples being graphene
and topological insulator surfaces.  In this paper we demonstrate that such systems can
host a remarkable variety of behaviors, in many cases controlled only by the density
of electrons in the system. Uniquely characteristic of these systems is an emergent
long-range form of the spin stiffness
when the Fermi energy $\mu$ resides
at a Dirac point, becoming
truly long-range as the magnetization density becomes very small.  It is demonstrated
that this leads to screened Coulomb-like interactions among domain walls, via a subtle
mechanism in which the topology of the Dirac electrons
plays a key role: the combination of
attraction due to bound in-gap states that the topology necessitates,
and repulsion due to scattering phase shifts, yields logarithmic interactions
over a range of length scales.  We present detailed results for the bound states in a particularly
rich system, a topological crystalline insulator surface with three degenerate Dirac
points and one energetically split off.  This system allows for distinct magnetic ground states which are
either two-fold or six-fold degenerate, with either short-range or emergent long-range
interactions among the spins in both cases.  Each of these regimes is accessible in principle by
tuning the surface electron density via a gate potential.  A study of the Chern number associated
with different magnetic ground states leads to predictions for the number of in-gap states that
different domain walls should host, which we demonstrate using numerical modeling are precisely
borne out. The non-analytic behavior of the stiffness on magnetization density is shown to have
a strong impact on the phase boundary of the system, and opens a pseudogap regime
within the magnetically-ordered region.  We thus find that the topological nature of
these systems, through its impact on domain wall excitations, leads to unique behaviors
distinguishing them markedly from their non-topological analogs.
\end{abstract}
\maketitle
\section{Introduction}
\label{sec:Introduction}
The study of magnetism hosted by dilute impurities in a non-magnetic metal has a long history in
physics, both for its fundamental interest and for possible applications such
systems might host.  The basic mechanism of magnetism in these systems was
first identified by Rutterman, Kittel, Kasuya, and Yosida \cite{RKKY1,RKKY2,RKKY3},
who demonstrated that magnetic impurity degrees of freedom can effectively couple
with one another through the conduction electrons.  Such ``RKKY interactions" between
two magnetic impurities
involves an induced, local spin polarization of the conduction electrons, due to
short range exchange interactions with an impurity spin.  The cloud of
induced spin density in the conduction electrons interacts with the second impurity some distance $R$ away,
so that the spin polarizations of the two impurities become effectively coupled.
This typically leads to an oscillating interaction with wavevector $2k_F$, with $k_F$ the
Fermi wavevector, contained in an envelope that falls off
as $1/R^2$ in two dimensions \cite{Kittel_1968,Fisher_1975}.
Viewed differently, in this mechanism the interaction between impurity spins is induced
by how they impact the total electronic energy of the conduction electrons, which is
sensitive to the relative orientation of the two spins \cite{Ketterson_book}.

Studies of RKKY interactions have enjoyed a significant resurgence in recent years, since the
advent of two dimensional electron systems with low energy dynamics controlled by a
Dirac equation.  Some examples include graphene, transition metal dichalcogenides,
and surfaces of various three-dimensional topological insulators.  These systems
host a variety of topological properties which impact the coupling among the
impurities as well as the types of magnetic states they host.  Perhaps the simplest example
is graphene \cite{Dugaev_2006,Brey_2007,Saremi_2007,Black_2010,
Fabritius_2010,Sherafati_2011,Kogan_2011,Lee_2012,Roslyak_2013,Gorman_2013,Crook_2015,Min_2017},
a two dimensional honeycomb lattice of carbon atoms,
for which the RKKY coupling between impurities $i$ and $j$
have a Heisenberg form (${\bf S}_i \cdot {\bf S}_j$), with equal magnitudes but of opposing sign
for impurity pairs on the same or opposite sublattices.  For doped graphene,
when the impurity density is sufficiently large compared
to $\pi k_F^2$, and quantum fluctuations are ignored, this leads to antiferromagnetic
order at zero temperature \cite{Brey_2007}.
The antiferromagnetism in this system is a consequence of the bipartite nature
of the graphene lattice, and contrasts with the ferromagnetic order expected in dilute magnetic semiconductors
\cite{Calderon_2002,Brey_2003}.
When the system is undoped, $k_F \rightarrow 0$ and the Fermi surface shrinks to two points,
leading to inter-spin coupling without oscillations and a faster
decay with distance ($1/R^3$).  Importantly, this $1/R^3$ behavior may be understood as
arising from non-analytic behavior in the static spin susceptibility of graphene at small wavevector {\bf Q},
which approaches its $Q=0$ value {linearly} with $Q$.  This behavior is actually rather generic for
electronic systems controlled by a Dirac Hamiltonian, and so applies to many systems of recent
interest beyond graphene.


Three dimensional topological insulators protected by time-reversal symmetry (TIs)
\cite{note_TI} offer an interesting related situation.
Because the bulk spectrum is gapped, electrons in the volume of the system are ineffective
at coupling spin impurities when the system is undoped.  However, the topological nature of the
band structure necessarily introduces
gapless states on their surfaces \cite{Hassan_2010,Qi_2011}.
As with graphene, impurity spins
exchange-coupled to the surface electrons develop effective inter-impurity interactions with a
long-range, monotonic character ($1/R^3$) when the Fermi surface is point-like.
Unlike graphene, this effective spin coupling is anisotropic due
to the strong spin-orbit
interactions typically present in these systems
\cite{Liu_2009,Biswas_2010,Garate_2010,Abanin_2011,Liu_Sau_2016}.
Depending on precisely
how the impurities couple to the surface electrons this is thought to lead to
ferromagnetic ordering or spin-glass behavior.  In the simplest cases,
a ferromagnetic groundstate should be stable, with the spin-anisotropic interaction
aligning the moments perpendicular to the surface.  From a mean-field perspective,
ferromagnetism is a natural outcome of the time-reversal symmetry-breaking it
entails, which gaps the surface spectrum and pushes the filled electron states
down in energy \cite{Efimkin_2014}.

Topological crystalline insulators (TCIs) \cite{Fu_2011,Liu_2013}
offer perhaps the richest of possible magnetic-dopant induced behaviors
among these systems \cite{Reja_2017}.  The paradigm of these are (Sn/Pb)Te
\cite{Hsieh_2012,Tanaka_2012,Xu_2012,Dziawa_2012,Zeljkovic_2015b}
and related
\cite{Okada_2013,Yan_2014,Zeljkovic_2015}  alloys.
The gapless surface states of these systems are protected by mirror symmetry \cite{Hsieh_2012},
so that generic breaking of time-reversal symmetry will not lead to lowering of
the electronic surface state energy per se \cite{Shen_2014,Assaf_2015}.  However, ferromagnetic
ordering with a spin component in the mirror plane breaks this symmetry, again gapping
the spectrum and pushing down the energies of filled electron states.  In most TCIs,
the crystal symmetry that protects the topology will dictate the presence
of more than one Dirac cone in the surface spectrum, and how this plays out depends
on the particular surface.  For example, topological (Sn/Pb)Te alloys host four Dirac
points for both (100) and (111) surfaces, but they are only fully degenerate in the
first case; in the second, one is energetically isolated while the remaining three are degenerate
(and related by three-fold rotations).  Because the system with such surfaces
has a variety of mirror planes, it can host
more than just the two-fold degenerate ferromagnetic groundstates found for
the TI surface: for a (100) surface one finds an eight-fold degenerate manifold
of ferromagnetic groundstates, while in the (111) case the system may be two-fold (Ising-like)
or six-fold degenerate \cite{Reja_2017}.  Moreover, in this latter case the system can be tuned to either
of the two types of ordering by controlling the surface electron density, in principle controllable
via an external gate.


An interesting aspect of the magnetically-doped TI and TCI systems is that they admit
low-energy topological excitations in the form of domain walls (DW's), linear regions
separating different possible groundstates of the system.
This is the subject of our study.
At low but
finite temperature, the energy per unit length of these structures controls how fast
the magnetization decays with temperature, and the loss of any net
magnetization above a critical temperature may be understood in terms of DW proliferation \cite{Jose_1977,Chaikin_1995}.
In typical ferromagnets, DW structure and energetics are determined by a balance of
the energetic cost of introducing gradients in the order parameter (favoring
wide DW's) and the energy associated with the magnetization failing to point
along a groundstate direction within the structure (favoring narrow DW's.)
Ignoring the effects of disorder in the impurity distribution, which throughout this
work we will assume in a coarse-grained model is qualitatively unimportant,
a simple continuum model for a
surface Dirac cone coupled to a surface magnetization ${\bf S}({\bf r})$
is a modified sine-Gordon model.  In writing this we assume
that a magnetization
perpendicular to the surface is favored (as for TI systems),
implementing the gap-opening effect of the magnetization.
The energy functional takes the form \cite{Rajaraman_book}
$E[{\bf S}]=E_2[{\bf S}] + E_g[{\bf S}]$,
where $E_2[{\bf S}] = -h\int d^2r S_z^2({\bf r})$
encodes the energetically-favored $\pm {\hat z}$ spin directions, and
the gradient energy $E_g$ is given by
$$E_g[{\bf S}] =
{{\rho_s} \over 2}\int d^2 r
\sum_{\mu,\nu=x,y}\sum_{i,j} \tilde{g}^{ij}_{\mu,\nu}
\partial_{\mu} S_i({\bf r}) \partial_{\nu} S_j({\bf r}).
$$
Here the constants
$\tilde{g}^{ij}_{\mu,\nu}$ encode anisotropy that descends from spin-orbit coupling in
the conduction electrons.  For a qualitative discussion we
assume $\tilde{g}_{\mu\nu}^{ij}=\delta_{ij}\delta_{\mu\nu}$.  In such a model, domain walls have an energy
per unit length $\varepsilon \sim \sqrt{\rho_s h}$ \cite{Rajaraman_book}.  The importance
of this energy scale shows up, for example, at the thermal disordering transition, where
from a balancing of entropy and energy \cite{Chaikin_1995} one expects
the transition temperature $k_B T_c \sim \varepsilon \ell$, where $\ell$
is a length scale over which the direction of the DW wanders, which typically is
the same as the DW width $\xi$.

In what follows we will argue that this energy estimate for DW's works well when
the Fermi energy cuts through the Dirac cones of the surface energy spectrum,
but fails when it aligns directly with a surface Dirac point.  The failure occurs
due to the simple form of the gradient energy $E_g$, which we will see is not
consistent with energetic estimates of the energy cost to introduce a gradient
in the spin.  Indeed this is anticipated by the
$1/R^3$ interaction form one finds in the perturbative RKKY analysis when the Fermi
energy is at a Dirac point.  Based on this one expects
a long-wavelength gradient functional of the form $E_g \rightarrow E_g^{LR}$, with
\begin{equation}
E_{g}^{LR}[{\bf S}] = {{\tilde{\rho}_s} \over 2}\int d^2 r_1 d^2 r_2
\sum_{\mu,\nu=x,y}\sum_{i,j} \tilde{g}^{ij}_{\mu,\nu}
\frac{\partial_{\mu} S_i({\bf r}_1) \partial_{\nu} S_j({\bf r}_2)}{|{\bf r}_1-{\bf r}_2|}.
\label{LRGrad_unscreened}
\end{equation}
This represents an
effectively {three}-dimensional Coulomb interaction among gradients on a two-dimensional plane.
Since DW's by their nature support a finite rotation of the magnetization, such a term will
lead to {\it logarithmic} interactions within and among the DW's.
In what follows, we will demonstrate that such long-range interactions do indeed appear
in these types of systems, albeit only up to a distance scale that diverges with vanishing
net magnetization.
In situations where the coupling between the magnetic impurities and conduction electrons
is small, this length scale can be quite large even in a magnetically ordered
situation. (For example, in graphene, for an exchange coupling $J \sim 5$meV \cite{Fritz_2013},
assuming a surface density of impurities per unit cell area $n_{imp}/a_0^2 = 4\%$,
it is of the order $(\hbar v_F/J)(a_0^2/n_{imp}) \sim 10\mu$m, where $v_F$
is the electron speed near the Dirac points.
Beyond this distance scale, we find that the gradient energy becomes
non-analytic in the amplitude of the magnetization.
This anomalous behavior presents itself both in systems where the
electronic states of two-component Dirac electrons have a spin-full character, and in
graphene, where there are separate Dirac spectra for each spin flavor.
The emergent long-range nature of the gradient energy impacts the DW energetics.
For example, the non-analytic behavior with magnetization amplitude at the longest wavelengths
should result
in DW energies that scale linearly with magnetization amplitude (adjustable
via the density of magnetic dopants).
In a course-grained theory, the spins appearing in the ${\bf S}_i \cdot {\bf S}_j$
coupling will each be proportional to spin density, leading to energies that are quadratic
in the magnetic impurity density for DW's in systems governed by short-range effective exchange interactions.
This should be reflected most directly in a critical temperature
for thermal disordering that scales {\it linearly} rather than
quadratically with impurity density, as we explain below.
%
In principle which of these behaviors is presented -- quadratic vs. linear in
impurity density -- may be chosen by
adjusting the density of conduction electrons on the surface,
either via a gate or by intentional doping.  Thus such magnets may be
tuned between rather different qualitative behaviors.

In systems where spin-orbit coupling is unimportant, such as graphene, the magnetic
degrees have a Heisenberg nature, and one does not expect DW's to form.
Indeed, these systems support gapless spin-wave modes
around the ground state
so that magnetic order will not set in at any finite temperature \cite{Mermin_1966}.
For short-range spin interactions these modes disperse linearly with wavevector \cite{Assa_book},
but if the stiffness changes to the long-range form above some wavevector scale, one
expects a crossover to $Q^{1/2}$ behavior.  Again, this crossover should occur only
in these systems when the Fermi energy is adjusted to be near the Dirac point energy,
allowing for in-principle tunable behavior.

The physics of DW's becomes even richer
in systems such as TCI's, in which
there are multiple
surface Dirac points.  In these systems the low-energy magnetization axis
is different for each Dirac point, leading to different possible numbers
of distinct ferromagnetic groundstate orientations.  For example, on the (111) surface
of materials in the (Sn/Pb)Te alloys, for an appropriately adjusted Fermi energy one
finds six degenerate groundstates \cite{Reja_2017}.  The low energy excitations
which connect these orientations are DW's.  Using numerical modeling which we present below,
one finds that the lowest energy of these connect orientations related by inversion
through the origin, followed by a 120$^\circ$
rotation around the normal to the surface.  In this way, the lowest energy
DW's connect all the different groundstate orientations into a six state clock model.
Thermal disordering in such a system should proceed in a two-step fashion,
in which long-range spin order is first lost as DW's proliferate, followed
by a vortex proliferation transition at higher temperature \cite{Jose_1977}.
Both transitions are believed to lie in the Kosterlitz-Thouless universality class.
As in the Ising case, we expect the emergent long-range interactions to impact
how the transition temperatures scale with impurity density,  and a
change in this behavior can in principle be
observed by adjusting the surface electron density.  Beyond this, a further
adjustment will bring the Fermi energy close to that of an energetically
isolated Dirac point, yielding two-fold degeneracy in the magnetization
groundstates, with either short-range or emergent long-range gradient energies
needed to model the DW energetics.  Thus we expect four distinct behaviors
for this surface, each accessible by adjusting the Fermi energy to an
appropriate value.  This is summarized in Fig. 1.
\begin{figure}[t]
\begin{center}
\includegraphics[width=1.2\linewidth, trim =50 0 -25 0, clip]{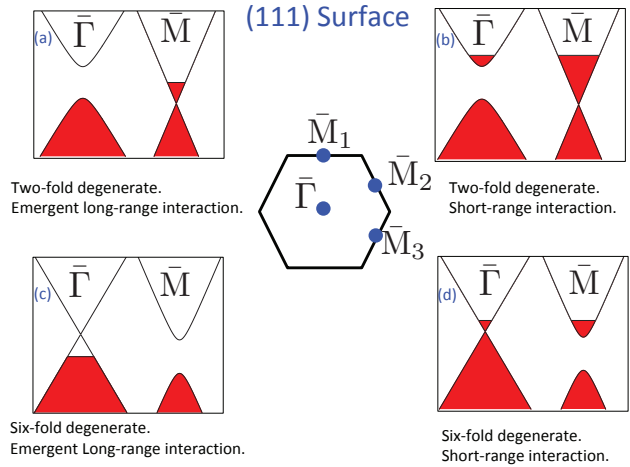}
\vspace{-1cm}
\caption{Summary of different magnetization behaviors expected for a (111) TCI surface.
The expected of dependence on $T_c$ on the magnetic impurity concentration $n_{imp}$ will vary
linearly or quadratically depending on whether there are short- or long- range spin-gradient
interactions in the system.  Inset: Locations of Dirac points in the surface Brillouin zone.
}
\label{fig:scheme}
\end{center}
\end{figure}

Another remarkable aspect of DW's in these systems are confined,
conducting states that they host \cite{Martin_2008,Liu_2009,Fang_2014,Jackiw_1976,Schaakel_2008}.
For a uniformly magnetized surface of
a TI or a TCI, symmetries broken by this (time-reversal in the former,
crystal symmetry in the latter) generically induce a Berry's curvature
in the vicinity of a surface Dirac point.  Importantly, when multiple
Dirac points are involved, this will occur for each in which the magnetization
opens a gap in the (local) energy spectrum.  We will see
explicitly for the concrete example of a TCI that integrating the
Berry's curvature in the vicinity of such points yields Chern
numbers $\pm 1/2$, so that the {\it change} in Chern number going
across the DW is always integral.  The numerical calculations
we present below demonstrate that one may understand the number
of conducting modes hosted by a given DW, as well as their chirality, from the change
in Chern numbers summed over all the Dirac points on the surface.

The presence of such conducting states in DW's opens unique opportunities
to interrogate them.  In principle they can be forced into a system
by pinning the direction of magnetization in opposite directions at
two ends of a sample at low temperature, or by quenching to low
temperature in zero magnetic field, freezing in thermally generated DW's.
The DW's could then be imaged, for example,
via STM spectroscopy on the surface, or detected indirectly by changes in the
surface conductivity due to their presence \cite{Dhochak_2015,Assaf_2015,Ueda_2015,Tian_2016}.
DW contributions to the dynamical conductivity might also be detected
via reflectance measurements from the surface. Such measurements could
also afford a window on thermal disordering of the surface magnetism,
at which point the DW's should proliferate.  While we expect the longest wavelength critical
fluctuations as one approaches thermal disordering to have a character
consistent with short-range gradient interactions \cite{note2,Sachdev_book,Vojta_2002},
there should exist a crossover regime in which the DW lengths and widths are impacted
by the emergent long-range interactions.  The existence of DW in-gap states thus introduces a
signal of the DW statistics that is measurable in probes coupling to the surface electrons.
In this way, domain walls
allow, in principle, direct access to the interesting physics that
emerges when magnetic degrees of freedom are introduced at TI and TCI surfaces.

This article is organized as follows.  We begin in Section II by
considering a simple Dirac electron model coupled to a static magnetization, and compute
the energy cost
coming from introducing gradients in the latter, with rather different behavior resulting
when the Fermi energy is at or away from the Dirac point.  A related analysis for graphene
is presented which yields results consistent with this, and we check this behavior numerically
to demonstrate that the physics remains valid in a tight-binding model.
We then turn in Section III to energetic calculations of DW {\it pairs}, in which we
demonstrate the presence of an emergent logarithmic interaction that appears as
the magnitude of the magnetization gets small.  Two analyses are presented.  The first involves
a transfer matrix method for a continuum model of Dirac electrons analyzed with
a phase shift method, where one finds that the behavior emerges from a near cancellation
of the DW separation dependence of the bound state energies, and the remaining spectral dependence found in phase shifts
of unbound electrons scattered by the DW's.  This is followed by a numerical analysis
of a tight-binding ``gapped graphene'' model that supports the result, demonstrating again
the consistency of continuum and microscopic models.  We then turn our attention to a more detailed study of
DW's in a TCI model in Section IV.
We begin with an outline of how we model these numerically, in particular explaining a technique
for projecting the Hilbert space into a set of surface states that allows us to focus
on the effects of magnetic moments near the surface.  We then apply this method to compute
the Berry's curvature and Chern numbers in the vicinity of surface Dirac points which become
gapped in the presence of a uniform magnetization.  This provides us with general expectations
for the number and chirality of states appearing in these gaps when there are DW's.
We then explain a method for numerically modeling DW's in this system, and present
results for several realizations of DW's.  In all cases we find that the number and
chirality of bound states within them are well-explained by the general expectations
arising from our Chern number calculations.  We also use this numerical method to demonstrate that in the six-state
TCI system, the lowest energy DW's are generically those that connect groundstates
that are closest in orientation.  This means that the system is best described as a
six-state clock model, rather than two sets of three states separated by a larger barrier.
Finally, in Section V we summarize our results, provide further discussion of their
significance, and possibilities for further exploration.

\section{Magnetization Gradient Energy}
\label{sec:gradients}

As discussed above, the unusual behavior of magnetic impurities coupled by Dirac electrons
is manifest when one introduces gradients in the magnetization.  In this section we demonstrate this
within two models of such systems.  The first is a simple model for electrons in a  surface system where
spin-orbit interactions are important, in which the electron wavefunctions involve two components, and
the electron spin degree of freedom is projected into these components.  These models arise
in the context of TI's and TCI's \cite{Efimkin_2014,Reja_2017}.  The second system we consider is graphene,
for which spin-orbit coupling is negligible.  The wavefunctions describe amplitudes for electrons
to be present on one of two sublattices of the carbon honeycomb structure, with either spin up or down,
and are thus four-component.  While in real systems the impurities are randomly located so that
disorder is present in the system, the relatively long-range of the effective spin-spin interactions when
$k_F$ is small or vanishing suggests one can coarse-grain the magnetization field over a large area so that
disorder effects become small, at least at long wavelengths \cite{Efimkin_2014,Reja_2017}.  For simplicity
we will ignore the effects of disorder in our analyses.

The underlying coupling between the impurity moments and the electron spin in these models is the $sd$ Hamiltonian,
$H_{sd} = J\sum_i {\bf S_i}\cdot {\bf s}({\bf r}_i)$, where ${\bf S}_i$ is a spin degree of freedom localized
at position ${\bf r}_i$, and ${\bf s}({\bf r})$ is the conduction electron spin field \cite{Yosida_book}.
These degrees of freedom may be deposited on the surface of the material, but for TI's and TCI's they
may be present in the bulk as well.  In the latter case, provided the Fermi energy of the system is in the bulk
gap, coupling among the bulk impurities will be exceedingly small, so that we expect them to be disordered
and for this reason negligible \cite{Rosenberg_2012,Lasia_2012}.  The spin impurities are however coupled near
the surface where conduction electrons are present.  Such models have the attractive feature that the
impurity atoms tend to enter as substitutional impurities at the same type of lattice site throughout
the crystal, so that there is considerable uniformity in the local coupling between spins and conduction
electrons \cite{Reja_2017}.

\subsection{Spin-Orbit Coupled Systems}

The coarse-graining approximation described above leads to a continuum form for the coupling Hamiltonian,
$H_{sd} \rightarrow \tilde{J}\int d^2 r {\bf S}({\bf r})\cdot {\bf s}({\bf r})$,  which then
must be projected into the low-energy sector of the electronic Hamiltonian.  The latter consists of
one or more single particle Dirac Hamiltonians, which with addition of the spin field ${\bf S}$ takes the
generic form
\begin{equation}
H = v_F\left\{(-i{{\partial} \over {\partial x}}-b_y) \sigma_1 + (-i{{\partial} \over {\partial y}}-b_x) \sigma_2 +  b_z \sigma_3 \right\},
\label{Hsurf_generic}
\end{equation}
where we have set $\hbar=1$, as we will throughout this paper, except where otherwise noted.
In this expression, $\sigma_i$, $i=1,2,3$ are the Pauli spin matrices, $v_F$ is the electron speed,
and the components of ${\bf  b}({\bf r})$ are proportional to projections of ${\bf S}({\bf r})$ onto
certain directions.  For example, for TI systems $b_3$ is proportional to the component of ${\bf S}$
perpendicular to surface \cite{Liu_2010,Silvestrov_2012,Efimkin_2014,Brey_2014}.  In (Sn/Pb)Te-type TCI systems, it is proportional to the spin
component along a particular $\Gamma$-$L$ direction in the bulk band structure \cite{Reja_2017}.
Note that more generally, the electron speeds along the $\hat{x}$ and $\hat{y}$ directions in the plane of the surface
may be different, but as this introduces no qualitative effects we ignore it for simplicity.

Our goal is to assess the cost in energy to the system when there is a spatial oscillation in ${\bf b}$
with some wavevector ${\bf Q}$, and we proceed to do this in perturbation theory.
For uniform ${\bf b}$, this Hamiltonian has the spectrum
$
\pm \varepsilon_0(q_x-b_y,q_y-b_x) = \pm v_F\sqrt{(q_x-b_y)^2 + (q_y-b_x)^2 + b_z^2}.
$
To this uniform ${\bf b}$ we add a
small oscillatory component $\delta {\bf b}$ with some definite wavevector ${\bf Q}$,
so that ${\bf b} = b_z \hat{z} + \delta {\bf b} \cos{{\bf Q} \cdot {\bf r}}$.  We then compute the
change in energy due to $\delta {\bf b}$ in perturbation theory, and examine its {\bf Q} dependence.
Shifting the origin of coordinates in momentum ($q_x'=q_x-b_2$, $q_y'=q_y-b_1$, with $b_x$ and $b_y$ the
in-plane components of the uniform {\bf b}-field) eliminates any effect of the uniform $b_{x,y}$ contributions.
The single-particle states diagonalizing Eq. \ref{Hsurf_generic} then have the form
\begin{equation}
|{\bf q},s\rangle = {1 \over {\sqrt{\Omega}}}
\frac{e^{i{\bf q} \cdot {\bf r}}}{\left[ q'^2 +(s\varepsilon_0(q')/v_F-b_z)^2 \right]^{1/2}}
\left(
\begin{array}{c}
   q_x'-iq_y'\\
   {s \over v_F}\varepsilon_0(q')-b_z
\end{array}
\right),
\label{spinor}
\end{equation}
where $\Omega$ is the surface area of the system, and $s =\pm 1$ labels the particle-
and hole-like states.

\begin{widetext}

\subsubsection{Fermi Energy in the Gap}
\label{EF0}
We first consider the situation where the Fermi energy is in the gap of unperturbed energy spectrum.
The change in the total energy of electrons is,
to leading non-vanishing order,
\begin{equation}
\Delta E = - \sum_{\bf q} \sum_{\bf p}
\frac{|\langle {\bf q},- | \delta h |{\bf p},+\rangle|^2}{\varepsilon_0({\bf q}) + \varepsilon_0({\bf p})},
\label{DeltaE}
\end{equation}
where $\delta h = \sum_{i=x,y,z}\delta b_i \cos({\bf Q} \cdot {\bf r}) \sigma_{i}$, with $\sigma_i$ the three
Pauli matrices, and we work in units for which
$v_F=1$.  Plugging into Eq. \ref{DeltaE} yields
\begin{equation}
\Delta E = - {1 \over 4} \sum_{\bf q} \left\{
\frac{|\langle {\bf q},- | \delta {\bf b} \cdot {\bf \sigma}|{\bf q}-{\bf Q},+\rangle|^2}{\varepsilon_0({\bf q}) +
\varepsilon_0({\bf q}-{\bf Q})} +
\frac{|\langle {\bf q},- | \delta {\bf b} \cdot {\bf \sigma} |{\bf q}+{\bf Q},+\rangle|^2}{\varepsilon_0({\bf q}) + \varepsilon_0({\bf q}+{\bf Q})} \right\}.
\label{DeltaEQ}
\end{equation}
Explicit calculations may be carried through with this expression, as we outline
in the Appendix.
To characterize the quadratic energy cost for magnetization
gradients we introduce a tensor quantity $g_{\mu\nu}^{ij}$
by
the definition  $\Delta E({\bf Q}) - \Delta E(0) =
{{\Omega} \over {2}}\sum_{\mu,\nu=x,y} \sum_{ij=x,y,z} g_{\mu\nu}^{ij}
Q_{\mu}Q_{\nu}\delta b_i \delta b_j$.
Many of the $g_{\mu\nu}^{ij}$ coefficients turn out to vanish; the non-vanishing ones
are given by
\begin{equation}
g_{xx}^{zz}=g_{yy}^{zz}=2b_z^2\int \frac{d^2q}{(2\pi)^2} \frac{q^2}{\varepsilon_0(q)^7}
={{8} \over {15\pi b_z}}
\label{g33}
\end{equation}
and
\begin{eqnarray}
g_{xx}^{xx}=g_{yy}^{yy} &=&
\frac{4}{5\pi b_z}, \nonumber\\
g_{xx}^{yy}=g_{yy}^{xx}=g_{xy}^{xy} &=&
\frac{16}{5\pi b_z}.
\label{g_in_plane}
\end{eqnarray}
An important property which must be checked is that the system is stable against gradients of the magnetization,
i.e., that the energy of the system can only increase as $Q$ increases from zero.  This is manifestly
true for gradients associated with $\delta b_z$.  For the in-plane components, it is convenient to notice
that one may write
\begin{equation}
\Delta E({\bf Q}) - \Delta E(0) = \left(
\begin{array}{cc}
  \delta b_x & \delta b_y
\end{array}
\right)
\left(
\begin{array}{cc}
  g^{xx}_{xx}Q_x^2 +g^{xx}_{yy}Q_y^2 &  g^{xy}_{xy}Q_xQ_y\nonumber \\
  g^{xy}_{xy}Q_yQ_x & g^{yy}_{xx}Q_x^2 +g^{yy}_{yy}Q_y^2 \nonumber
\end{array}
\right)
\left(
\begin{array}{c}
\delta b_x \nonumber \\
\delta b_y
\end{array}
\right).
\label{MatrixEnergy}
\end{equation}
Using Eqs. \ref{g_in_plane}, it is easy to confirm that
the eigenvalues of the matrix appearing in this equation are always positive for any direction of
${\bf Q}$, and increase quadratically with its magnitude.  This indicates that gradients in
the magnetization tend to increase the energy of the configuration, so that the spin-spin interactions
favor ferromagnetism in this system.

A prominent feature of these results is that all these coefficients diverge as the gap-opening
component $b_z \rightarrow 0$,
indicating a diverging stiffness as the uniform component of the surface magnetization vanishes.  On the other hand, if the oscillations in the
underlying ${\bf S}$ field come from rotations in the field, but the field itself is of
constant length, then we expect $\delta {\bf b}$, $b_z$ $\sim |{\bf S}|$, so that $\Delta E({\bf Q}) -\Delta E(0)$
is still non-analytic in ${\bf  S}$ and is anomalously large when $|{\bf S}|$ is small,
but is not divergent in the ${\bf S} \rightarrow 0$ limit.

This surprising
result is actually consistent with the effective RKKY spin coupling that is known for graphene;
as discussed in the introduction, the $1/R^3$ interaction found there leads to long-range gradient
interactions, with a Fourier transform that is {\it linear} rather than quadratic in $Q$, and
hence non-analytic in wavevector.  Our perturbative calculation explicitly assumes that $\Delta E({\bf Q})$
is analytic in wavevector, and the divergence of the stiffnesses as $b_z \rightarrow 0$ is the
signal that this assumption breaks down. To see more clearly how this works, we will consider
the energetic cost of imposing a spin gradient on electrons in graphene.  Before proceeding with
this, however, we extend the analysis discussed above to the case where the electron system is
doped, and see that this relieves the large gradient energy found in the calculation above.

\subsubsection{Fermi Energy in a Band}
When the Fermi energy $\mu$ is alternatively in the band, we end up with a very different result:
there is no dependence on the wavevector {\bf Q} to order $Q^2$. Again
the perturbation around a uniformly magnetized state will take the form
\begin{equation}
\delta h =
\delta {\bf b} \cdot \vec{\sigma} \cos {{\bf Q} \cdot {\bf r}}.
\label{delh}
\end{equation}
In what follows we assume the Fermi energy $\mu$ is in the valence band -- i.e., below the gap.
Because the Hamiltonian is particle-hole symmetric we should obtain the same result for
$\mu \rightarrow -\mu$.
Assuming $Q < \mu$,
the change in energy due to the perturbation can be expressed
at second order as a sum of two terms, $\Delta E = \Delta E_+ + \Delta E_-$, with
\begin{equation}\label{DeltaEpl}
\Delta E_+ =
{1 \over 4} \sum_{\mathclap{\substack{q>k_F \\
|{\bf q} -{\bf Q}|<k_F}}}
\frac{|\langle {\bf q},- | \delta {\bf b} \cdot \vec{\sigma}
|{\bf q}+{\bf Q},-\rangle|^2}{\varepsilon_0({\bf q}+{\bf Q}) - \varepsilon_0({\bf q})}
-{1 \over 4} \sum_{q>k_F}
\frac{|\langle {\bf q},- | \delta {\bf b} \cdot \vec{\sigma}
|{\bf q}+{\bf Q},+\rangle|^2}{\varepsilon_0({\bf q}+{\bf Q}) + \varepsilon_0({\bf q})},
\end{equation}
where the Fermi wavevector is defined by $\varepsilon_0(k_F)=\mu$.
$\Delta E_-$ has the same form as Eq. \ref{DeltaEpl}, with ${\bf Q} \rightarrow -{\bf Q}$.
As demonstrated in the Appendix, when the $\Delta E_+$ and $\Delta E_-$ are
summed, the result is {\it independent} of ${\bf Q}$;
i.e., the energy required to introduce an oscillation
in the magnetization is independent of the oscillation wavevector.
This indicates that an effective energy functional for the
magnetization should have vanishing coefficient for the quadratic
gradient term -- effectively, a vanishing spin stiffness.  This
contrasts dramatically with the situation we found for $\mu=0$,
where the stiffness diverged as $b_z \rightarrow 0$.

Two comments are in order.  The first is that this vanishing stiffness results from
the perfect linear spectrum of our unperturbed model.  In real systems there
is some curvature in the spectrum away from the Dirac point energy, and we
expect this will lead to non-vanishing contributions to the stiffness.  If
the Fermi energy is not too far from the Dirac point then one can treat such
deviations perturbatively, and these should be finite.  Thus we expect non-vanishing
contributions for spin gradients in a doped system, as will be supported by our
numerical studies described below, but these will be small compared to what happens
when the Fermi energy is in the gap of the uniformly magnetized system.  The second
is the comparison of this result to a closely related one for graphene: when doped, its spin
susceptibility is independent of ${\bf Q}$ for small $Q$ \cite{Brey_2007}.
In this situation, however, RKKY interactions between spins do {\it not} vanish, due
to contributions from large $Q$.  This leads to ferromagnetic coupling among spins
on the same sublattice, and antiferromagnetic ones for spins on opposite sublattices,
for length scales shorter than $\sim 1/2 k_F$ \cite{Brey_2007}.  Beyond this scale,
the RKKY interactions oscillate and average to zero.  The net effect is a short distance coupling,
which ultimately leads to a non-vanishing gradient energy for the system..

As we see, the comparison of this system with the behavior of graphene is quite useful, so
we next turn to an analysis of what happens in the latter system when a spin gradient is imposed.


\end{widetext}

\subsection{Comparison to Graphene}

Because graphene has essentially no spin-orbit coupling, it couples
to an impurity spin in a different way than what was examined in the
last section.  Nevertheless, results for it do bring some
insight to systems governed by the Hamiltonian $H$ appearing
in Eq. \ref{Hsurf_generic}.  In graphene
the spin operator is completely independent of the spinor degree
of freedom that $H$ acts upon; spin is a separate
quantum number for the electrons.  The effect of a single impurity
spin is to act like a local Zeeman field with direction fixed
by the impurity spin itself.

\subsubsection{Perturbation Theory}

In the standard perturbative approach to RKKY interactions \cite{Ketterson_book},
one computes the static linear spin response
$\chi^{ij}_{\alpha \beta}(Q)$ of (the Fourier transform of)
the electron spin components $s_i({\bf Q})$ to a perturbation
$JS_j({\bf Q})$, where $J$ is the $sd$ coupling and
$\alpha,\beta$ = A,B are indices specifying the
sublattice(s) to which the impurities are coupled.
The spin symmetry dictates that the spin response has the
form $\chi^{ij}_{\alpha \beta}(Q)= \chi_{\alpha\beta}^0 \delta_{ij}$,
and the total change of energy at second order in $J$
is $\Delta E = -J^2\sum_{\bf Q} \sum_{i=x,y,z} \sum_{\alpha,\beta}
\chi^0_{\alpha\beta}(Q)S_{i,\alpha}(-{\bf Q})S_{i,\beta}({\bf Q})$,
where $S_{i,\alpha}$ is the $i$th component of the impurity spin field
on sublattice $\alpha$.

As has been shown previously \cite{Brey_2007}, for undoped graphene $\chi_{\alpha\beta}^0(Q)$
begins at a positive cutoff-dependent constant for $Q=0$ and varies
{\it linearly} with increasing $Q$: for example,
$\chi_{AA}^0(Q)={1 \over {4\pi}}\left(\Lambda-{\pi \over 8}Q\right)$,
where $\Lambda$ is an upper cutoff of order the bandwidth.
For doped graphene
$\chi_{AA}^0(Q)$ is independent of $Q$ (and equal to the
$Q=0$ value for the undoped case) up to $Q=2k_F$, where a non-vanishing
slope in $Q$ sets in.
($\chi_{AB}$ has the same magnitude as $\chi_{AA}$ but has opposite
sign.) The cusp is a realization of the well-known Kohn anomaly
and leads to $2k_F$ oscillations in the response.

The results are reminiscent of what we found in the last two
subsections.  The linear behavior in $Q$ for undoped graphene is
non-analytic and indicates that the quadratic small $Q$ calculation
carried out above must fail in the limit that the gap closes
i.e., for vanishing uniform magnetization in the zero-doped, spin-orbit
coupled model.   Indeed we expect that for ${b_z} \rightarrow 0$ that the spin-response
associated with Eq. \ref{Hsurf_generic} will tend to a combination
of $\chi_{AA}^0$ and $\chi_{AB}^0$ for graphene.  Thus, we should understand
the divergences in Section \ref{EF0} in this limit as
indicating a crossover from quadratic to linear behavior in the
spin response with respect to $Q$ when the system exits the
broken symmetry state.


\subsubsection{Beyond Perturbation Theory: Helicity Modulus}
\label{sec:graphene_stiffness}

In contrast to the models considered above, in graphene the expected ordering
at low temperature is antiferromagnetic across the sublattices \cite{Brey_2007}.
When this is present the RKKY interaction as calculated perturbatively
fails at the longest length scales in a way very analogous to what
happened in the spin-orbit coupled case.  This occurs
because a uniform staggered magnetization acts as a mass term in
the Hamiltonian for each spin individually, opening a gap $\Delta$ in the spectrum.
If one works perturbatively around this state, one expects an exponential
falloff in the spin-spin interaction at length scales beyond that set by $\Delta$.
Interestingly, since spin-orbit coupling is essentially negligible in this system,
no spin orientation is favored, and it is possible to assess the energetics of
spin gradients of different length scales, as we now show.

Suppose the {\it staggered} magnetization is characterized by an ordering vector ${\bf b}({\bf r})$.
For a square system of linear size $L$ one can imagine a configuration in which ${\bf b}$
rotates precisely once around some fixed axis as ${\bf r}$ varies down the entire
length of the sample in some direction.
The helicity modulus \cite{Chaikin_1995} is defined in terms of the energy cost to introduce
this spin twist, relative to a uniform groundstate:
\begin{equation}
\rho_s(g=2\pi/L) = \lim_{L \rightarrow \infty} 2L^{-2}\left[E(g=2\pi/L)-E(g=0)\right] /(2\pi)^2,
\label{helicity_def}
\end{equation}
where $g$ is the wavevector of the imposed spin gradient,
and $E(g)$ is the energy of the system (proportional to its area) with some imposed spin gradient.
While $\rho_s(g=0)$ is the spin stiffness of the system at the longest possible length scale
available in a finite size system, we can generalize this quantity by allowing $g$ to
be a free variable, probing the energy cost for gradients at length scales $2\pi/g$.
This quantity may be computed for graphene subject to a uniformly rotating staggered
magnetization.

Our Hamiltonian in this situation is
\begin{equation}
H_G=v_F\left[ \hat{p}_x\tau_x +\hat{p}_y \tau_y - {\bf  b} \cdot {\vec \sigma} \tau_z \right],
\label{G_Ham}
\end{equation}
where $\vec\sigma$ is the set of Pauli matrices acting on the spin degree of freedom,
$\vec\tau$ are the corresponding matrices acting in the sublattice space, and $\hat{p}_{x,y}$
are components of the momentum operator.  As above
we set $v_F=1$.  If
${\bf b} = {b}_0(\sin\theta,0,\cos\theta)$ is independent of position, then the eigenstates of
${\bf b}\cdot\vec{\sigma}$ are
\begin{align}
\chi_+=
\left(
\begin{array}{c}
\cos\frac{\theta}{2} \nonumber\\
\sin\frac{\theta}{2}
\end{array}
\right),
&\quad
\chi_-=
\left(
\begin{array}{c}
-\sin\frac{\theta}{2} \nonumber\\
\cos\frac{\theta}{2}
\end{array}
\right),
\label{spinors}
\end{align}
and the corresponding eigenergies of $H_G$ are given by $\pm \sqrt{p_x^2+p_y^2+b_0^2}$,
each of which is two-fold degenerate.  To compute $\rho_s(g)$ we will need to find the single-particle
energies in the situation where $\theta \rightarrow g x$.  To do this we transform our
spin quantization axis to be {\it locally} parallel to ${\bf b}(x)$.  This is equivalent
to writing the eigenstates of Eq. \ref{G_Ham} in the form
\begin{equation}
\Psi(x) = {\mathbf \alpha}(x) \otimes {\bf \chi}_+(x) + {\mathbf \beta}(x) \otimes {\bf \chi}_-(x),
\label{Psi}
\end{equation}
where the $\sigma$ matrices act on the vectors ${\bf \chi}_{\pm}$, and the $\tau$ matrices
act on the two-component vectors ${\mathbf \alpha}$ and ${\mathbf \beta}$.  With some algebra, one can show that the
stationary state equation $H_G\Psi=\varepsilon\Psi$ can be cast in the form
\begin{equation}
(\tilde{H}-\varepsilon)
\left(
\begin{array}{c}
{\mathbf \alpha}\\
\mathbf\beta
\end{array}
\right)
=0,
\label{Eval_tildeH}
\end{equation}
where
\begin{equation}
\tilde{H}
=
p_x\tau_x +p_y\tau_y -b_0\tau_z\mu_x-g\mu_z\tau_x,
\label{final_H}
\end{equation}
and the $\vec{\mu}$ Pauli matrices act in the $(\alpha,\beta)$ space.

\begin{figure}[thbp]
\includegraphics[width=0.9\linewidth]{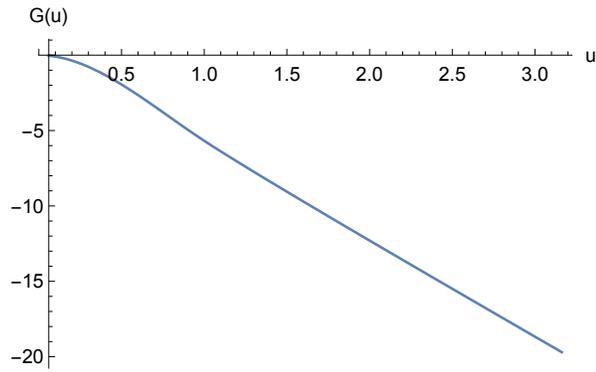}
\caption{Numerical result for $G(u)$ (Eq. \ref{BigIntegral}) as a function of $u$.
}
\label{G}
\end{figure}

The solutions to Eq. \ref{Eval_tildeH} can be evaluated directly, yielding four single
particle energies,
\begin{equation}
\pm{\varepsilon}_s({\bf p})= \pm \left\{
p^2+g^2+b_0^2 + 2sg\sqrt{p_x^2+b_0^2}
\right\}^{1/2},
\label{energies}
\end{equation}
where $s=\pm 1$.  We are interested in the situation where the negative energy
states are completely full, so the total energy is
\begin{equation}
E(g)=-\sum_s\sum_{{\bf p}}{\varepsilon}_s({\bf p}).
\label{total energy}
\end{equation}
From this we wish to subtract the energy at $g=0$.  The single particle energies
of the filled states are clearly $-\sqrt{p^2+b_0^2} \equiv -\varepsilon_0({\bf p})$.
The energy difference $E(g)-E(0)$ can be written in the form
\begin{widetext}
\begin{equation}
\Delta E(g) \equiv E(g)-E(0) = -\sum_p \left[\varepsilon_{+1}({\bf p}) + \varepsilon_{-1}({\bf p})
-\varepsilon_0({\bf p} - g\hat{x}) - \varepsilon_0({\bf p} + g\hat{x})
\right].
\label{energy_diff}
\end{equation}
The shift of the $g=0$ energies in the subtraction does not affect the result provided
the system obeys periodic boundary conditions, and in this form one may confirm that
the sum over ${\bf p}$ in
Eq. \ref{energy_diff} is independent of cutoff.  Substitution yields the explicit expression
\begin{eqnarray}
\Delta E(g) &=& -\sum_{{\bf p}} \Biggl\{
\left[p^2+b_0^2+2g\sqrt{p_x^2+b_0^2}+g^2\right]^{1/2}
+\left[p^2+b_0^2-2g\sqrt{p_x^2+b_0^2}+g^2\right]^{1/2} \nonumber \\
&-&\left[p^2+b_0^2+2g p_x+g^2\right]^{1/2}
-\left[p^2+b_0^2-2g p_x+g^2\right]^{1/2}
\Biggr\}.
\label{Delta_E}
\end{eqnarray}
\end{widetext}
Assuming the system to be of sizes $L_x$ and $L_y$ in the $\hat{x}$ and $\hat{y}$
directions respectively
we can replace the momentum sum in the thermodynamic limit by an integral.
If we assume $g << b_0$, to lowest non-trivial order in $g$ we find
\begin{equation}
\Delta E (g) \approx L_xL_yg^2b_0^2 \int \frac{d^2p}{(2\pi)^2}\frac{1}{\varepsilon_0(p)^3} \sim L_xL_yg^2b_0.
\label{small_g_scaling}
\end{equation}
The result is anomalous in the sense that, for a generic magnet where the stiffness usually
depends analytically on the magnetization scale, we expect $\rho_s \sim L_xL_y\Delta E(g)
\sim b_0^2g^2$.  Eq. \ref{small_g_scaling} is consistent with a long-range interaction
among spin gradients that is cut off by the scale of the magnetization itself, $b_0$.
This interpretation is further supported by considering larger values of $g$.  To do
this, we compute the $p_y$ integral in $\Delta E(g)$ analytically, which allows it
to be cast in the form $\Delta E(g) = -L_xL_y \frac{g^3}{(2\pi)^2} G(\frac{b_0}{g})$,
with
\begin{widetext}
\begin{eqnarray}
G(u) \equiv \int_{-\infty}^{\infty} &dx&
\Biggl\{
\left[1+u^2+x^2\right]
\log\left[
\frac{(1+u^2+x^2)^2-4x^2}{(1-u^2-x^2)^2} \right] \nonumber \\
&+&2x \log\left[
\frac{1+2x+x^2 + u^2}{1+2x+x^2 + u^2} \right] \nonumber \\
&-& 2\sqrt{x^2+u^2}
\log \left|\frac{1+\sqrt{x^2+u^2}}{1-\sqrt{x^2+u^2}} \right|
\Biggr\}.
\label{BigIntegral}
\end{eqnarray}
\end{widetext}
Note in writing this expression we have taken the momentum cutoff to infinity.
One may compute $G(u)$ numerically, with the result that $G(u) \sim -u^2$
for $u<<1$, and $G(u) \sim -|u|$ for $u>>1$, as illustrated in Fig. \ref{G}.
The latter result reproduces
the explicit small $g$ result, while the former shows $\Delta E(g) \sim b_0^2 |g|$
for $b_0 << g$.  This non-analytic behavior in $g$ is what one expects from the linear
$Q$ behavior of the spin susceptibility discussed in the previous
subsection, indicative of long-range interaction for magnetization gradients.
We see however that the interaction is cutoff by the average magnetization $b_0$.
This length scale can become very large in the limit of low magnetic impurity density or
a relatively small $sd$ coupling scale $J$.

\begin{widetext}

\begin{figure}[t]
\includegraphics[width=0.9\linewidth]{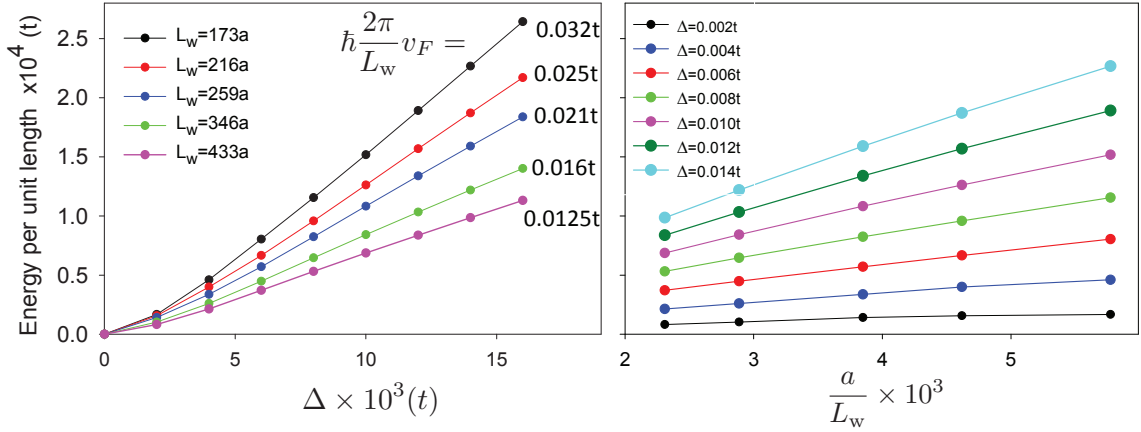}
\vspace{-.8cm}
\caption{Numerical calculation of energy per unit length required for single overturn of the staggered
magnetization $\Delta$ in a graphene ribbon of width $L_w$, relative to that with a uniform
staggered magnetization, for various values of $L_w$ and $\Delta$.
When $L_w$ is held fixed, this energy grows linearly with $\Delta$.  For fixed $\Delta$, the
energy approaches a constant as $1/L_w$ grows.}
\label{stiff_ribbon}
\end{figure}

\end{widetext}

The result holds as well for graphene when treated in the tight-binding model.
To show this, we consider the simplest such system in which the carbon atoms
are laid out in a triangular lattice with two atoms per unit cell and lattice
parameter $a$, with nearest neighbor hopping $t$.
The Fermi velocity is related to the tight-binding parameters via $\hbar v_F = \sqrt{3}/2 t a_0$.
In each unit cell there is an effective Zeeman energy ${\bf  h} = \Delta \hat{z}$, in opposite
directions for each of the sublattices, modeling the staggered magnetization.
We consider a ribbon of this, with cross-sectional width $L_w$, in which ${\bf  h}$
rotates around an axis by $2\pi$
along the ribbon cross-section. The system has well-defined momentum along the
$\hat{y}$-direction, $p_y$, and for each of these we compute a set of single-particle
energies by diagonalizing the tight-binding model numerically.  The relevant
electronic energy of the system
is the sum of all negative energy states, integrated (numerically) over $p_y$.
From this we subtract the corresponding energy for a uniform staggered magnetization
${\bf h}$, with the same magnitude $\Delta$.  This difference is $\Delta E(g=2\pi/L_w)$.
When $v_F/L_w \gg \Delta$,  one expects $\Delta E \sim L_wg$ becomes constant as $L_w$
grows.  By contrast, for fixed $L_w$ it should grow linearly with increasing $\Delta$.
This behavior is consistent with the numerical observations,
as illustrated in Fig. \ref{stiff_ribbon}.

\subsection{Discussion}

We conclude this section with some observations as well as speculations
regarding the impact of the unusual gradient energy
in these systems.  In the context of DW's, one interesting consequence
is how the energetics impacts the temperature at which the system
should disorder.  A simple estimate \cite{Vanderzande_book,Shankar_book}
of the free energy to create a DW of length $L$ against an otherwise uniform
magnetization background takes the form
$\Delta F(L)= \varepsilon L-k_B T \eta (L/\xi)$, where $\varepsilon \sim \sqrt{\rho_s\bar{S}}$
is the energy per unit length, with $\bar{S}$ the average magnetization per unit
area, and $\xi \sim \sqrt {\rho_s/\bar{S}}$  is the width of a DW, $T$ is the
temperature, and $\eta$ is factor of order unity which characterizes how quickly
the DW changes its direction as one moves down its length, in units of $\xi$;
the second (entropic) term arises from the number of configurations one may
construct for the DW, in which the complicating factors of interactions among
different parts of the DW have been ignored, as well as the fact that a finite
$L$ DW in a system without boundaries is actually a closed loop.  In spite of
these simplifications, for the Ising model the condition $\Delta F(L) <0$,
which is interpreted as DW proliferation and the loss of magnetic order in the system,
yields an estimate of $k_BT_c=\varepsilon\xi/a \sim \rho_s$.  In the Ising model,
this type of argument yields the correct $T_c$ to within 25\% of the exact answer \cite{Shankar_book}.

In the present system, however, the behavior of $\rho_s$ is anomalous.  For example,
for short-range interactions this scales as $\rho_s \sim \bar{S}^2$, which in
turn is proportional to the {\it square} of the impurity density, since the $\rho_s$
is a long-wavelength measure of interactions among the impurities.  If one uses the
long wavelength estimate for systems analyzed above, we find $\rho_s \sim \bar{S}$,
{\it linearly} proportional to the impurity density.  This behavior contrasts
with what happens when the Fermi energy is moved away from any Dirac point energy
of the surface, in which case we return to a magnetic system with short-range
interactions: $T_c$ then scales quadratically with impurity density.
This change in behavior is an in-principle measurable signature of the interesting
DW energetics in these systems.

In addition to the anomalous average magnetization dependence of DW's in this system,
our gradient analysis suggests an emergent long-range interaction which becomes
important at increasingly long length-scales as the magnetization decreases.
In the next section, we will demonstrate the presence of this interaction
by examining the energetics of inter-DW interactions.

\section{Domain Wall Interactions}

As discussed above, one aspect of the unusual gradient interactions in these systems
would be the emergence of long-range interactions between DW's as the magnetization
scale gets small.  To test this, we will compute these interactions directly in
two simple models: continuum Dirac electrons coupled to a piecewise constant magnetization
field, and a tight-binding model of ``gapped graphene.''
In both cases we will see that the character
of the interaction changes significantly depending on the placement of the chemical
potential $\mu$: when it passes through the magnetization-induced gap, it becomes
increasingly long-ranged as the magnetization becomes small.  When $\mu$ is outside
this gap, the interaction remains short-ranged even as the gap closes.

\subsection{Continuum System with Piecewise Constant Magnetization: Phase Shift Analysis}

We begin with a generic surface Dirac Hamiltonian of the form in Eq. \ref{Hsurf_generic},
which within regions of constant magnetization may be written as
\begin{equation}
H=(k_y-b_y)\sigma_x+(k_x-b_x)\sigma_y+\Delta\sigma_z,
\label{eq_ham_TM}
\end{equation}
where $k_x$ and $k_y$ are components of the electron wavevector for the system surface
with constant magnetization.  In this equation we have taken our unit of energy
to be $\hbar v_F/a_0$, where $v_F$ is the speed associated with the Dirac point when
$\Delta=0$, and our length unit $a_0$ is set by a microscopic lattice scale.
Our approach will be to consider linear combinations
of the eigenstates associated with this type of Hamiltonian, matching them across boundaries
where $b_x$, $b_y$, and $\Delta$ change suddenly.  We compute a transfer matrix for the
system, from which we can obtain both bound state energies and phase shifts for scattered
states, allowing us to compute the energies of each of these as a function of separation
between two DW's.  We will see the effects of these combine in a surprising way to yield
a slow variation of the system energy when the chemical potential is in the gap, and the
separation is not too large.
\begin{figure}[!b]
  \begin{center}
    \includegraphics[width=0.4\textwidth]{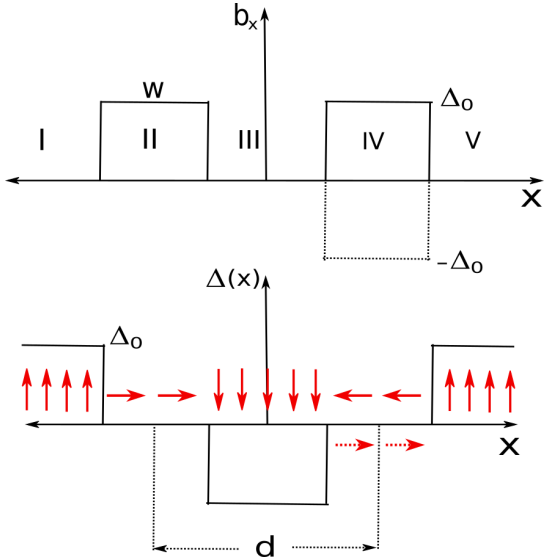}
  \end{center}
\caption{\rm Piecewise constant domain wall configuration with two domain walls of width $w$ each separated by
a center-to-center distance $d$.  Top panel: The parameter $b_x$, for which we allow the
possibility of values that are equal or opposite, allowing for magnetizations that rotate with the
same or with opposite senses.  Piecewise constant regions $I - V$ are labeled.
Bottom panel: $\Delta(x)$, illustrating that the magnetization rotates
from up to down and back again.  Arrows indicate the orientation of the magnetization vector
in each region.}
\label{fig_piecewise}
\end{figure}
%


\subsubsection{Wavefunctions}

The type of DW configuations we analyze are illustrated in Fig. \ref{fig_piecewise},
which contain five separate regions in which $\Delta$ and $b_x$ are constant, labeled $I$ through $V$.
The rotation of the magnetization within the two DW's may have the same or opposite senses,
as illustrated by the solid and dashed arrows in Region $IV$.
We treat this as a scattering problem where
electrons
in regions $I$ and $V$ are connected through a transfer matrix $T$,
\begin{equation}
\vec{\Psi}_V=T\vec{\Psi}_I
\label{eq_transfer}
\end{equation}
The two components of the wavefunctions $\vec{\Psi}_i$ represent amplitudes for the two orbitals
upon which the Dirac matrices in Eq. \ref{eq_ham_TM} act.
To obtain the transfer matrix $T$, we find eigenvectors of this Hamiltonian for some fixed energy $E$ in each region,
and match both components of the wave functions at each boundary ($I$ to $II$, $II$ to $III$, etc.)
Note that $k_y$ is a good quantum number and is constant for a wavefunction in all regions.

The general form for the wavefunction in region $j$ may be written as
\begin{equation}
\vec{\Psi}_j=e^{ik_yy}\left[ A_je^{i(k_x^{+})_jx}\left(\begin{array}{c} u_{+}^j \\ v_{+}^j \end{array}\right)+B_je^{i(k_x^{-})_jx}\left(\begin{array}{c} u_{-}^j \\ v_{-}^j \end{array}\right)\right],
\label{eq_wave_fun}
\end{equation}
%
and
energy $E=-\sqrt{k_y^2+k_x^2+\Delta_0^2}$ the same in all regions.  We solve this
straightforwardly to obtain the values of $k_x$ for the scattering states in
regions $I$ and $V$; note for bound states this may turn out be imaginary.  The
energy also
determines the values of $(k_x^{\pm})_j$ in each of the ``internal'' regions $j=II,III,IV$,
\begin{equation}
(k_x^{\pm})_j=-b_x^j\pm\sqrt{E^2-k_y^2-\Delta_j^2}.
\end{equation}
In terms of these the
values of $u,v$ are given by
\begin{eqnarray}
u_{\pm}^j=\frac{k_y-i((k_x^{\pm})_j+b_x^j)}{\sqrt{(\Delta_j-E)^2+k_y^2+((k_x^{\pm})_j+b_x^j)^2}},\nonumber\\
v_{\pm}^j=\frac{E-\Delta_j}{\sqrt{(\Delta_j-E)^2+k_y^2+((k_x^{\pm})_j+b_x^j)^2}}.\nonumber
\end{eqnarray}
With this information, the $T$ matrix may be straightforwardly computed analytically;
the expression is lengthy and we do not present it explicitly.
Note that the $T$ matrix contains the information about the domain wall width $w$ and the distance $d$ between the two domain walls (DWs).  We will compute the energy of the DW structure from $T$, which contains
two contributions of similar size, one from bound states induced by the DW's, and one from
scattering phase shifts.

\subsubsection{Energy from Bound States}
We can express the scattering amplitudes in terms of the components of $T$ using
\begin{equation}
A_V=T_{AA}A_I+T_{AB}B_I, B_V=T_{BA}A_I+T_{BB}B_I,
\end{equation}
where $A$ and $B$ are the amplitudes for right- and left- moving electrons,
respectively, in the $I$ and $V$ regions. To obtained the bound state solution, we put $k_x\rightarrow i\kappa$
and find $\kappa$ such that $A_I=B_V=0$. This condition is satisfied when $T_{BB}=0$.

For a given $w$ and $d$, we numerically find the solution for $\kappa$ which gives $T_{BB}=0$.
A very nice simplification for this particular geometry is that the
solution is independent of $k_y$, which
makes the computation of this energy contribution particularly simple, once $\kappa$ is known.
Since the model we are considering is particle-hole symmetric, we need only consider
chemical potentials $\mu \le 0$.  The
total energy contribution from the bound states is
given by summing over all states with energy below $\mu$, which includes only negative energy
states,
\begin{equation}
\Delta E_b/L_y=-\frac{1}{\pi}\int_{k_y^c}^{\pi}dk_y\sqrt{k_y^2+\Delta_0^2-\kappa^2},
\label{BS_integral}
\end{equation}
where $L_y$ is the length of the system along the $\hat{y}$ direction.  Note the
lower cutoff $k_y^c$, which is given by
\[
    k_y^c =
    \begin{cases}
       \sqrt{\mu^2-\Delta_0^2+\kappa^2}  & \text{if $(\mu^2-\Delta_0^2+\kappa^2)\geq 0$,}\\
        0 & \text{otherwise,}
    \end{cases}
\]
is non-trivial because a bound state will only be present below a given $\mu$ if $k_y$ is sufficiently large.
The integral in Eq. \ref{BS_integral} is straightforward to compute with the numerically
generated values of $\kappa$.

\subsubsection{Energy from Scattering States}
We next need to calculate the change in electronic energy due
phase shifts of the wave functions due to scattering from the DWs.
To do this we imagine the whole system to be embedded at the center of a large box of length
$L$, whose size we will eventually take to infinity.  For simplicity we
require the lower component of the wave function to vanish at the edges of this
box. (Other boundary conditions may be considered but should not have qualitative effects on the results.)
Using Eqs. \ref{eq_transfer} and \ref{eq_wave_fun}, this leads to a condition for the allowed states in this box,
\begin{equation}
\frac{T_{AA}e^{ik_xL}-T_{AB}}{T_{BB}e^{-ik_xL}-T_{BA}}=1.
\end{equation}

This may be rewritten as a quadratic equation for $e^{ik_xL}$ in terms of the matrix elements of $T$, whose
solutions we cast in the form $e^{ik_xL}=e^{iq_nL+i\eta_{\pm}(k_x)} \equiv e^{ik'L}$. The values of $\eta_{\pm}(k_x)$ give
allowed values of $k_xL$, with the $\eta_+$ solutions corresponding to $q_n \equiv n\pi/L$, with $n$ the even
positive integers, when the DWs are eliminated ($T \rightarrow$ unity), and with $\eta_-$ corresponding
to $q_n$ with $n$ odd in the same limit.
Interestingly, we again find a useful independence from $k_y$: for a given $k_x$ the total phase
shift $\eta(k_x)=\eta_{+}(k_x)+\eta_{-}(k_x)$ is independent of $k_y$.
The shift in energy due to the DW structure comes from the differences between
$q_n$ and $k'(q_n)
$, which, though small, add up to a finite contribution
when summed over all the occupied states. To see this one starts with the expression
for the total energy contribution due to the scattering states,
$$
E_{ph}/L_y= \int {{dk_y} \over {2\pi}} \sum_{n,\,filled} E[k_x(q_n),k_y].
$$
For large $L$, we recast the sum over $n$ as a momentum integral,
$$
E_{ph}^{\pm}/L_y\rightarrow \int {{dk_y} \over {2\pi}} \int {{Ldq} \over {2\pi}}E[k_x(q),k_y].
$$
Here $\pm$ corresponds to the two solutions for the phase shift, $\eta_{\pm}(k_x)$. Now using the
relation $k_xL=q_nL+\eta_{\pm}(k_x)$, the energy may be written as
$$
E_{ph}^{\pm}/L_y=\int {{dk_y} \over {2\pi}} \int \frac{dk_x}{2\pi}(L-\frac{d\eta_\pm(k_x)}{dk_x})E[k_x,k_y].
$$

The first term gives a constant background which is independent of the DW
separation, and so maybe ignored. Adding the non-trivial contributions from $\eta_+$ and $\eta_-$,
we obtain the energy increase due to scattering,
$$
\frac{\Delta E_{ph}}{L_y}=-\frac{1}{4\pi^2}\int \int dk_ydk_x\frac{d\eta(k_x)}{dk_x}E[k_x,k_y].
$$
As mentioned above, the total phase shift $\eta(k_x)$
is independence of $k_y$, so we may rewrite the above equation as
$$
\frac{\Delta E_{ph}}{L_y}=-\frac{1}{4\pi^2}\int dk_x\frac{d\eta(k_x)}{dk_x}\int dk_yE[k_x,k_y].
$$
Note that the domain of integration for $k_x,k_y$ must respect the condition $E[k_x,k_y]<-|\mu|$.
When the chemical potential is in the gap for the uniformly magnetized system, both $k_x$ and $k_y$ will
vary from $-\pi/a_0$ to $\pi/a_0$ for some cutoff scale $\Lambda = \pi/a_0$.

Since our analysis above yields explicit expressions for $\eta(k_x)$ (again, not presented as this
is lengthy yet straightforward to obtain), it is convenient to integrate this directly rather
than its derivative.  Up to surface terms which are independent of the DW separation,
partial integration yields
\begin{equation}
\frac{\Delta E_{ph}}{L_y}=-\frac{1}{2\pi^2}\int_0^{\pi/a_0}dk_x\eta(k_x)\frac{dF(k_x)}{dk_x}
\label{PS_integral}
\end{equation}
where $F(k_x)$ is given by
\begin{equation}
F(k_x)=\int_{k_y^c}^{\pi/a_0}dk_y\sqrt{k_x^2+k_y^2+\Delta_0^2}.
\end{equation}
The lower limit $k_y^c$ is again defined as
\[
    k_y^c =
    \begin{cases}
       \sqrt{\mu^2-\Delta_0^2-k_x^2}  & \text{if $(\mu^2-\Delta_0^2-k_x^2)\geq 0$,}\\
        0 & \text{otherwise.}
    \end{cases}
\]
The integral in Eq. \ref{PS_integral} is straightforward to evaluate numerically.

\subsubsection{Results}

\begin{figure}[!b]
  \begin{center}
    \includegraphics[width=0.8\linewidth]{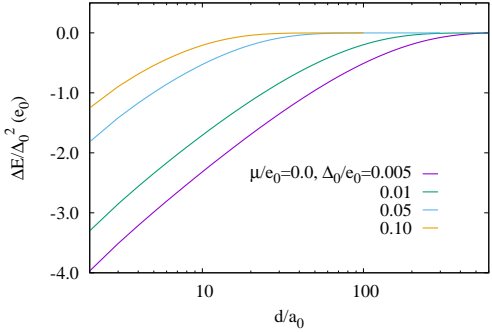}
  \end{center}
  \vspace{-0.8cm}
  \caption{DW pair energy (adding the contributions from both the bound state and phase shift) as a function of the distance between the domain walls $d$ when $\mu$ is in the gap. Different lines indicate
different values of $\Delta_0$, as indicated.  Energies expressed in units of $e_0 \equiv \hbar v_F/a_0$.}
  \label{fig_long_range_a}
\end{figure}
\begin{figure}[!b]
  \begin{center}
    \includegraphics[width=0.8\linewidth]{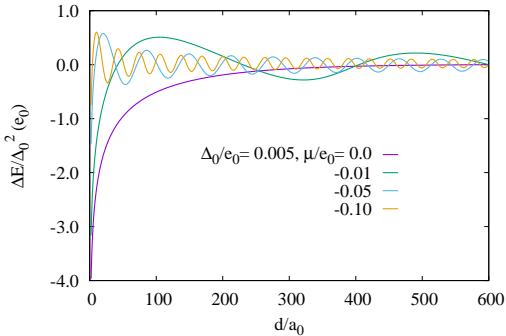}
  \end{center}
  \vspace{-0.8cm}
  \caption{DW pair energy (adding the contributions from both the bound state and phase shift) as a function of the distance between the domain walls $d$ when $\Delta_0=0.005$ for various values of $\mu$,
as indicated. Energies expressed in units of $e_0 \equiv \hbar v_F/a_0$.}
  \label{fig_long_range_b}
\end{figure}

We next turn to a discussion of results from this analysis. In all cases the basic energy
scale is set by the square of the gap energy $\Delta_0$, which we scale out in presenting
the results.  Distances are shown in units of the cutoff length scale $a_0$, which may be taken
for concreteness as the lattice constant of the underlying structure.
Fig. \ref{fig_long_range_a} illustrates typical results for the energy of a pair of DW's as a function of their
separation, for different values of the gap $\Delta_0$ for the uniform magnetization far from the pair,
when the chemical potential $\mu$ is in the gap.
In these calculations, the DW widths are taken to be 0
so that the magnetization jumps discontinuously
at each DW.  The results are shown on a linear-log scale, and it is apparent for the smallest values
of $\Delta_0$ that the energy rises nearly linearly towards the asymptotic value for well-separated
DW's.  This behavior is expected for interactions between spin-gradients that vary as $1/R$, so that
the interaction between line-like objects such as a DW will be logarithmic.  As expected
from our analysis above, this long-range
interaction is emergent, in the sense that it is cut-off at a distance scale that diverges as
$\Delta_0$ vanishes.  We find very similar results for finite width DW's, for both cases
where the in-plane spins are parallel or antiparallel (see Fig. \ref{fig_piecewise}.)
The basic interaction between DW's is set by the change in gap-opening component of
the field, not the components perpendicular to this.

\begin{figure}
  \begin{center}
    \includegraphics[width=0.99\linewidth]{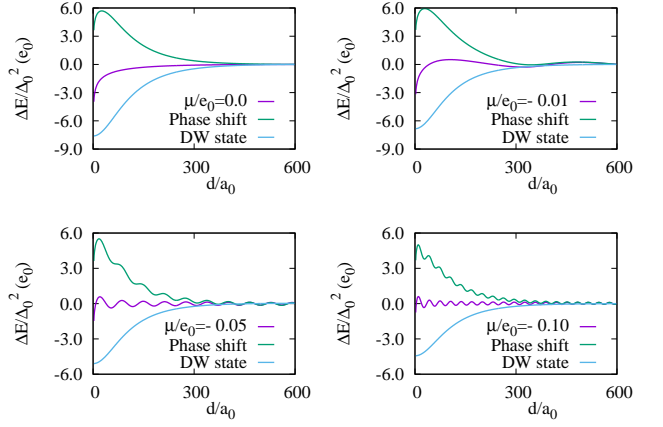}
  \caption{Contributions to DW pair energy from bound states and phase shifts shown separately, for $\Delta_0=0.005 e_0$ and for various values of chemical potential as indicated.
Energies expressed in units of $e_0 \equiv \hbar v_F/a_0$. The near cancellation
of the two contributions is apparent.
  }
\label{fig_comparison_fig}
    \end{center}
\end{figure}

Fig. \ref{fig_long_range_b} illustrates corresponding results for fixed $\Delta_0=0.005$,
in units of $\hbar v_F/a_0$,
for different values of $\mu$.
Here it makes most sense
to present the results on a linear scale, and it is apparent that effective range of the
DW attraction shrinks as $\mu$ moves deeper into a band.  The expected $2k_F$ oscillations
are also apparent.  Figs. \ref{fig_long_range_a} and \ref{fig_long_range_b} firmly
establish the qualitative differences between DW interactions for $\mu$ in a gap and
$\mu$ in a band.

\begin{figure}[!b]
  \begin{center}
    \includegraphics[width=0.99\linewidth]{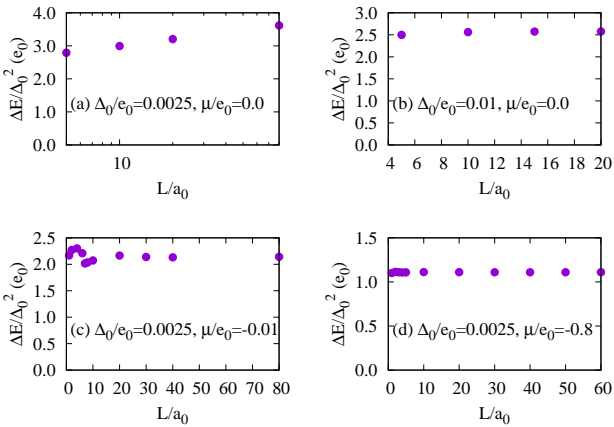}
  \end{center}
  \vspace{-1cm}
  \caption{ Energy of DW superlattice as a function of separation $L$ for graphene, for different
  values of $\mu$ and $\Delta_0$ as indicated, in energy units of $e_0 \equiv \hbar v_F/a_0$.
 Domain wall are one unit cell wide.
  Note x-axis is on a log scale in (a), but on a linear scale in the other figures.}
  \label{graphene_dw_ints}
\end{figure}

As discussed above, these interactions arise from the combined effects of the bound states
in the DW's and the phase shifts of the scattering states.

It is interesting to examine the
contributions of these separately, as we do in
Fig. \ref{fig_comparison_fig}.
Interestingly,
one finds an attractive bound state contribution which slightly overbalances a repulsive
phase shift contribution, to yield a net attractive interaction.  The ranges of each
individually turn out to be considerably longer range than the net attraction, and their sums
yield the characteristic behaviors illustrated in Figs. \ref{fig_long_range_a} and
\ref{fig_long_range_b}.  This is a surprisingly intricate way for the slow $d$ dependence
of the interaction that emerges at small $\Delta_0$ to be realized microscopically:
our expectations of its presence descended from perturbative analyses
around uniform magnetized systems, which contain no obvious signals that the DW's
will host bound states at all.  This behavior is a remarkable demonstration of
how the topological character of the underlying bands -- which necessitate
the presence of the bound states -- plays a powerful if subtle role in
yielding the long-wavelength physics of the magnetic degrees of freedom
in this system.


\subsection{A Microscopic Realization: Gapped Graphene}

The results in the previous subsection were derived in the context of a continuum model
with an imposed short length-scale cutoff.  To further establish the presence of the
emergent long-range interaction, we wish to see that it is present in a microscopic,
i.e., a tight-binding, model.  To do this, we consider a model of spinless
electrons in a graphene lattice, with a staggered potential that varies in the
$\hat{x}$ direction.  In general, in such a staggered potential graphene is a normal insulator;
however, under certain circumstances it does have a non-trivial topological character.
This behavior emerges because
each valley carries a half integer Chern number of opposite sign.  In geometries for which
valleys are not admixed, the system will behave in ways akin to more protected topological
systems. For example, when there are regions of opposing staggered potential  $\Delta_0$
meeting at a valley-preserving interface,
valley-dependent gapless chiral modes are known to emerge \cite{Xiao_2007,Yao_2009}.

The staggered potential we employ in our model has four regions: one with amplitude
$\Delta_0$, one with amplitude $-\Delta_0$, separated by two regions where the
staggered potential vanishes for one unit cell along the $\hat{x}$ direction.
These two regions are a distance $L$ apart, and model DW's in the system.
The entire system obeys periodic boundary conditions along the $\hat{y}$ direction,
and is periodic in the $\hat{x}$ direction up to a phase $e^{ik_x(2L+2a)}$, with $a$ the DW width (equivalent to
the basic unit cell size in our model). The system may be understood as a superlattice
of DW's, with the total number of DW pairs given by the number of $k_x$ values
retained in the calculation.  A corresponding wavevector $k_y$ for the $\hat{y}$-direction
is also a good quantum number, and the number of $k_y$ values retained effectively fixes
the size of the system in this direction.  Finally, the microscopic lattice structure
is oriented such that the centers of the two valleys (${\bf K}$ and ${\bf K}'$ points)
are separated along the $\hat{y}$ direction, avoiding valley-mixing effects
\cite{Brey_2006,Palacios_2010}.

To assess the energetics of this system, we compute the total electronic energy for negative
energy states up to some choice of chemical potential $\mu$,
and subtract from this the
corresponding energy for a system of the same size with uniform staggered magnetization $\Delta_0$.
$\mu$ may be chosen to be in the gap or within a band of the latter.  Note that the spectrum
is particle-hole symmetric, so we only examine non-positive values of $\mu$.
This energy difference is a measure of the energy required to create the DW pairs,
and by varying $L$ we obtain a measure of their interaction energy.

Fig. \ref{graphene_dw_ints} illustrates some typical results.
In panel (a) we illustrate the
DW pair energy as a function of $L$, on a linear-log scale, for a small value of $\Delta_0$
and $\mu$ in the gap.  The straightness of the line
clearly attests to the logarithmic interaction in this distance scale.
For large enough $L$ we expect the interaction energy to reach a
constant value, and this behavior is demonstrated in panel (b) for
larger $\Delta_0$, where the asymptotic length scale is not so large that it is difficult
to reach numerically.  Panels (c) and (d) contrast these with the situation for $\mu$ in a
band, where it is clear that the interaction is much shorter in range.  Note that the
$2k_F$ oscillations are not apparent in these figures; this is due to the
number of ${\bf k}$ values retained (20 $k_x$ values, 1001 $k_y$ values) which leads
to a relatively small number of bands cutting through the chemical potential.  In principle
a much larger number of $k_x$ values should bring out the oscillations, but in practice
we find this requires a smaller number of $k_y$ values, which we find sacrifices
accuracy at short distances.  Thus, although these numerics are limited by the absence
of the expected $2k_F$ oscillations at long distances, they do confirm the transition
from logarithmic to short-range behavior (for small $\Delta_0$) as $\mu$ moves into a
band.

\begin{widetext}

\begin{figure}
  \centering
    \includegraphics[width=0.75\textwidth]{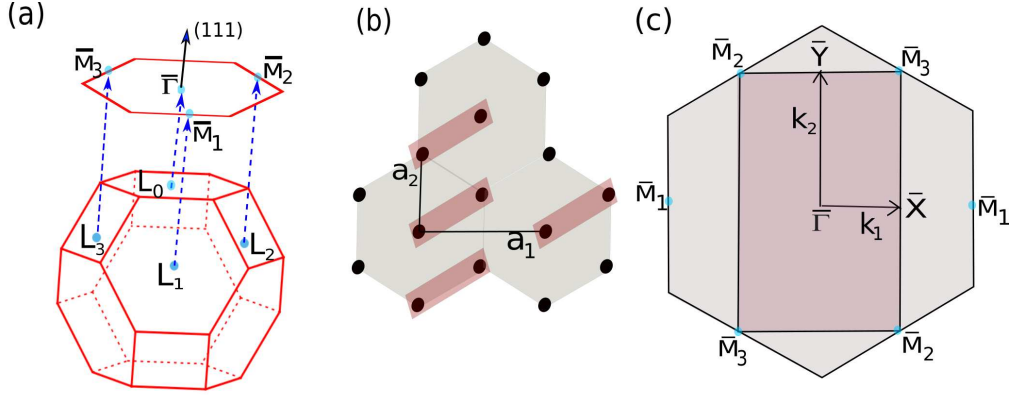}
    \caption{(a) The fcc Brillouin zone containing $L_i (i=0,1,2,3)$ points and their projections
onto the (111) surface, which
    yield the
    $\bar{\Gamma}$ and ${\bar{M}_i} (i=1,2,3)$ points. (b) Extended real space unit cells with two atoms per
     unit cell, used in constructing a domain wall.  (c) Surface Brillouin zone for the two surface atom
     unit cell, which folds the original hexagonal Brillouin zone for the single atom (real space) unit cell
     into a rectangular one.}
    \label{fig_bz}
\end{figure}

\end{widetext}

\section{Domain Walls in TCI Materials}

As discussed above, interactions among DW's in Dirac-mediated systems
involves a delicate balance of the energetics
of the bound states they host and the scattering of unbound states.
Moreover, the possibility of detecting the DW's is greatly enhanced
by the bound states because they render the DW's conducting.
While the analyses discussed above have largely focused on magnetic
moments at a surface coupled by a single Dirac point, many systems
actually host multiple points, all coupling to the magnetic moments
and contributing to the effective interactions among spin gradients.  In this last section,
we study this in some detail for the interesting case of TCI materials,
where the competition among these can lead to multiple orientations for
the ground state energy \cite{Reja_2017}.  In particular we will demonstrate
that for the (111) surface of TCI's in the (Sn/Pb)Te class, for a uniform
magnetized system each distinct Dirac point has an associated Chern number
of $\pm 1/2$, and that the total change of Chern number across a DW correctly
predicts the number of states hosted, independent of details of the DW
structure.  We will also present numerical evidence that the DW energetics
strongly suggest that these systems should be described by a six-state
model under appropriate circumstances.

\subsection{Tight-Binding Model}

TCI's such as (Pb/Sn)Te have band topology protected by
mirror symmetry. The Bravais lattice of the system
is fcc with two sublattices (i.e, a rocksalt structure), which we
label $a$ and $b$. Focusing on the (111) surfaces, it is convenient
to view the structure as two-dimensional triangular lattices with
ABC stacking.  In this orientation,
triangular layers of $a$ and $b$ atoms are arranged alternately along the (111) direction.

A ``standard'' tight-binding model for these systems
is given by \cite{Littlewood_2010,Hsieh_2012}
$H_{bulk} = H_m+H_{nn}+H_{nnn}+H_{so}$, with
\begin{widetext}
\begin{eqnarray}
H_m &=&\sum_j m_j \sum_{{\bf R},s}
{\bf c}^{\,\,\dag}_{j,s}({\bf R}) \cdot {\bf c}_{j,s}({\bf R}) ,
\quad\quad\quad\quad \nonumber \\
H_{nn} &=& t \sum_{({\bf R},{\bf R^{\prime}}),s}
{\bf c}^{\,\,\dag}_{a,s}({\bf R}) \cdot {\bf d}_{{\bf R},{\bf R^{\prime}}}
{\bf d}_{{\bf R},{\bf R^{\prime}}} \cdot {\bf c}_{b,s}({\bf R^{\prime}}) + h.c. , \nonumber\\
H_{nnn} &=& \sum_j t_j^{\prime} \sum_{(({\bf R},{\bf R^{\prime}})),s}
{\bf c}^{\,\,\dag}_{j,s}({\bf R}) \cdot {\bf d}_{{\bf R},{\bf R^{\prime}}}
{\bf d}_{{\bf R},{\bf R^{\prime}}} \cdot {\bf c}_{j,s}({\bf R^{\prime}}) + h.c. , \nonumber\\
H_{so} &=&
i \sum_j \lambda_j \sum_{{\bf R},s,s'} {\bf c}^{\,\,\dag}_{j,s}({\bf R}) \times
{\bf c}_{j,s^{\prime}}({\bf R}) \cdot (\vec \sigma)_{s,s^{\prime}}. \quad\quad
\label{eq_ham}
\end{eqnarray}
\end{widetext}
In these equations
${\bf R}$ labels the sites of a cubic lattice, $j=a,b$ are the species type
(Sn/Pb or Te),
which have on-site energies $m_{a,b}$,
and $s = \uparrow,\downarrow$ is the electron spin.  The 3-vector of operators
${\bf c}_{j,s}({\bf R})$ annihilates electrons in $p_x$, $p_y$ and $p_z$ orbitals,
and there is a local spin-orbit coupling strength $\lambda_j$ on each site.
($\vec \sigma$ is the vector of Pauli matrices.)
The vectors ${\bf d}_{{\bf R},{\bf R^{\prime}}}$ are unit vectors pointing from
${\bf R}$ and ${\bf R^{\prime}}$, and, finally, the sum over $({\bf R},{\bf R^{\prime}})$
denotes positions which are nearest neighbors, while
$(({\bf R},{\bf R^{\prime}}))$ denotes next nearest
neighbors.  The bulk energy structure of these systems includes direct
energy gaps in the vicinity of $L$ points of the Brillouin zone \cite{Littlewood_2010,Hsieh_2012},
whose locations are illustrated in Fig. \ref{fig_bz}(a).  There are four such (distinct) points,
located on hexagonal faces of the Brillouin zone, and there is a three-fold
rotational symmetry around each $\Gamma-L$ axis.

To focus on surfaces, we will consider slab geometries of this system, to which
we will add magnetic moments.  In the absence of any magnetization,
the system hosts gapless surface states \cite{Liu_2013} whose energies
are within the bulk gap.  These states form the ``low-energy sector''
in which we are interested, and which ultimately control the coupling
of magnetic moments near the surface.  In these materials magnetic dopants
may be added throughout the bulk \cite{Inoue_1975,Inoue_1977,Inoue_1979,Story_1986,Karczewski_1992,
Geist_1996,Geist_1997,Prinz_1999,Lusakowski_2002},
which typically substitute for atoms at the (Sn/Pb) sites.
The doping also introduces carriers in the bulk (moving the chemical potential
out of the gap), creating RKKY coupling among the bulk magnetic moments.
The system in this way becomes a dilute magnetic semiconductor.  The model
we consider \cite{Reja_2017} supposes that compensating dopants can be added to the system
to remove the bulk electrons, bringing the chemical potential back to the bulk gap,
and eliminating any significant coupling among
the bulk magnetic moments.  This effectively eliminates these degrees of freedom
on average \cite{Rosenberg_2012,Lasia_2012}.  Conducting electrons at the system boundary however will
still be present due to their topological protection, so that magnetic moments
near the surface form an effective two-dimensional magnet.  These are the
degrees of freedom upon which we wish to focus.

The calculations we describe below begin with a slab with 47
layers, which we find to be sufficient to avoid significant mixing between states on the two
surfaces. The tight-binding parameters we use
in Eq. \ref{eq_ham} are adapted from Ref. \onlinecite{Fulga_2016}, and
are specifically (using the nearest neighbor hopping $t$ as our energy
unit) 
$t_a^{'}=-t_b^{'}=-0.556t,
\lambda_a=\lambda_b=-0.778t, m_a=-m_b=3.889t$.
The simplest unit cell for our slab geometry incorporates
one site from each
triangular layer, so that our system is effectively a two-dimensional triangular
lattice with many atoms in the unit cell.   The resulting surface Brillouin Zone (BZ) is a
hexagon, which is perpendicular to one of the $\Gamma$-$L$ directions
as shown in Fig. \ref{fig_bz} (a).  We denote this particular $L$-point as $L_0$,
and its projection onto the surface BZ is denoted as $\bar{\Gamma}$.  The projections
of the other three $L$ points are denoted as $\bar{M}$ points in the surface BZ.

The large unit cell and orbital basis for our model in principle allows a full band structure
calculation for the slab geometry, but produces a very large number of bands, most of which are
far in from the ``low-energy'' part of the spectrum.  Incorporation of all these bands severely
limits the realizations of DW's we can in practice consider in the slab.  Moreover, for
the Chern number calculations we describe below, fully including all of these introduces
large numerical errors.  To circumvent these problems, we project our system into
a Hilbert space that incorporates the surface states, i.e., those states with energy within or closest in energy
to the center of the bulk band gap.

\subsection{Chern Number}

We begin by demonstrating numerically that the Chern number associated with each surface Dirac
point is $\pm 1/2$.  To do this, we adopt a method detailed in the Ref. \onlinecite{Fukui_2005}.
Briefly, the method involves discretizing the momentum space within the surface BZ, computing phases
associated with each plaquette in the discretized space which become equivalent to the local
Berry's curvature when the discretization becomes sufficiently fine, and summing over these
to obtain a Chern number.  The phases can be defined for every band, allowing a computation
of the Chern number for each of them.

In practice, when there are many bands these calculations become numerically
difficult.
The challenge arises because in
regions where different bands approach the Berry's curvature varies rapidly, and one needs a very
fine $k$-space mesh to resolve this with sufficient accuracy.  For large unit cells such as the
slab we consider, such calculations are impractical.  For narrower slabs the computations
can be carried through, but only for such narrow ones that the states on the two surfaces are
strongly admixed.  As we are interested in Chern numbers for individual surfaces, we instead
project the Hilbert space of the wide-slab system into the set of bands that host surface states,
and examine their Berry's curvature directly.

\begin{figure}
  \centering
    \includegraphics[width=0.8\linewidth,trim= 0 -30 0 0 ,clip]{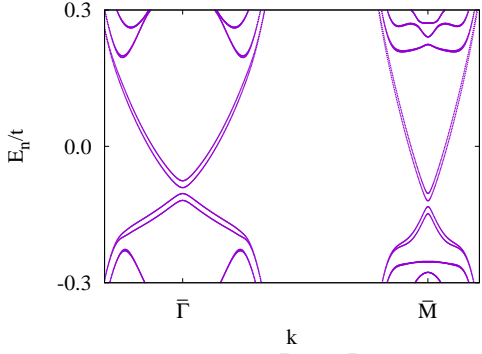}
    \vspace{-1cm}
    \caption{The band structure around $\bar{\Gamma}$ and $\bar{M}$ with magnetic moment $|J{\bf S}|=0.05$. A small potential gradient has been introduced to lift the surface degeneracy.
}
    \label{fig_band_structure}
\end{figure}

The bands associated with surface Dirac cones only develop well-defined Chern numbers when they are gapped
out, and we are interested specifically in what these are for the uniform magnetized states
that are connected by a DW.  We thus carry out our calculations for the slab system, with
uniform magnetic moments ${\bf S}$ at the $(Pb/Sn)$ sites, coupled to the electrons via an $sd$ Hamiltonian,
$\sum_i J{\bf S} \cdot {\bf s}_i$, where ${\bf s}_i$ is the electron spin at site $i$
at a surface.  Here ${\bf S}$ for each surface
points along
the $\Gamma$-$L_0$ axis, which maximizes the gap opening of the Dirac point at the
$\bar{\Gamma}$ point.
We then
focus on the two bands that host the top and bottom surface Dirac cones.
These two bands are well
separated in energy from other bands around symmetry points ($\bar{\Gamma}, \bar{M}$) as shown in Fig. \ref{fig_band_structure}, but come very close to the bulk
bands as they enter the bulk spectrum. This makes it very difficult to
calculate the Berry's curvature accurately
too far away from the $\bar{\Gamma}$ and $\bar{M}$ points in the surface BZ \cite{Fukui_2005}.

To proceed we assume that the Berry's curvature away from the
symmetry points ($\bar{\Gamma}, \bar{M}$) summed over all the bands with energies below the
center of the gap average
to zero, and focus on the contributions from the surface bands. To identify these individually
for each surface, we break
the symmetry between the top and bottem surfaces of the slab by adding a very small
potential gradient.  As shown in Fig. \ref{fig_band_structure}, this separates out the two
surface bands and allows us to follow them individually.

\begin{figure}
  \begin{center}
   \includegraphics[width=0.90\linewidth]{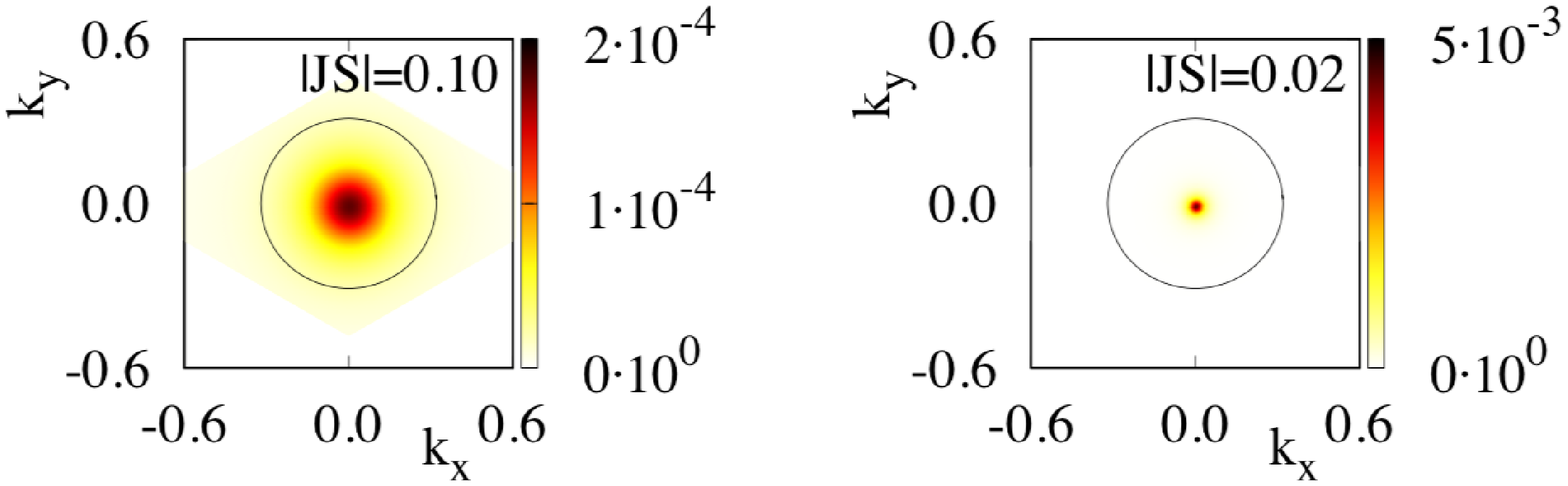}
    \caption{The Berry's curvature for top surface state around $\bar{\Gamma}$ point with the magnetic moment
    $|J{\bf S}|$ as indicated.}
    \label{fig_berry_gamma}
  \end{center}
\end{figure}

\begin{figure}
\begin{center}
    \includegraphics[width=0.99\linewidth]{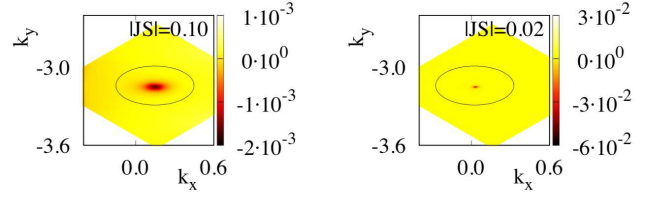}
    \caption{The Berry's curvature for top surface state around $\bar{M}$ point with the magnetic moment
    $|J{\bf S}|$ as indicated.}
    \label{fig_berry_M}
\end{center}
\end{figure}

\begin{figure}
\begin{center}
    \includegraphics[width=0.8\linewidth]{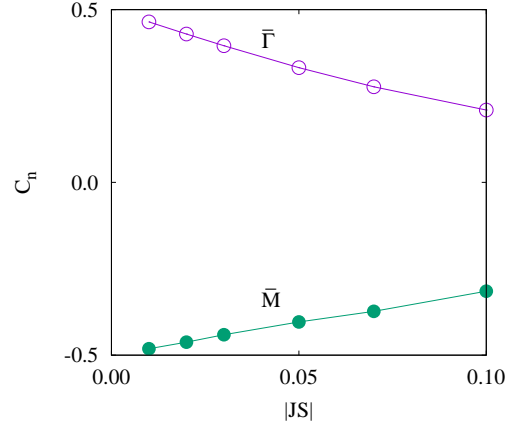}
    \caption{The extrapolation of Chern number with magnetic moment $|J{\bf S}|$ for top surface state.
    The bottom surface state has opposite behavior i.e., the values are opposite in sign.}
    \label{fig_chern}
\end{center}
\end{figure}

Fig. \ref{fig_berry_gamma} shows our computed Berry's curvature for the top surface state around $\bar{\Gamma}$ point
for $|J{\bf S}|=0.10$ and  $0.02$. It is evident that the most of the curvature accumulates around the
symmetry point, which becomes more localized with decreasing magnetization strength $|J{\bf S}|$.
We then calculate
the Chern number by numerically
integrating the  curvature within a circle outside of which the curvature is very small,
as indicated in Fig. \ref{fig_berry_gamma}.  The ``leakage'' of Berry's curvature outside
this circle becomes increasingly negligible as $|J{\bf S}|$ becomes small, and we find that
as
$|J{\bf S}|\rightarrow 0$, the Chern number tends to
$1/2$ as shown in Fig. \ref{fig_chern}. Similar behavior
occurs around the $\bar{M}$ points.  The Berry's curvature illustrated in Fig. \ref{fig_berry_M}
clearly becomes more localized with decreasing magnetization, and
the extrapolated integrated Berry's curvature
tends to $-1/2$, as
shown in Fig. \ref{fig_chern}.
Note that for the opposite surface, for magnetizations pointing outward at both surfaces,
the Chern numbers for the Dirac spectra at the same type of symmetry point have {\it opposite}
sign.
This can be understood as a consequence of a combination of time-reversal and inversion symmetries
(in the absence of the imposed potential gradient), which map states on each surface onto one another.

These results have important consequences for DW's, which connect regions with different uniform
magnetizations.  The change in Chern number topologically necessitates the presence of chiral, conducting bound states
within a DW, with chirality given by the sign of that change \cite{Hatsugai_1993}.
For example, in the Ising case, where a DW connects states of magnetization
parallel and antiparallel to the surface,
one expects 1 and 3 states, of opposite chirality, for the $\bar{\Gamma}$ and $\bar{M}$ points,
respectively.  We now turn to numerical
investigations that show this to be the case, and that it holds robustly with respect to
parameters that characterize the details of the DW structure, as to be expected for
a topologically protected property.

\subsection{Domain Walls on a TCI Surface}
We now turn to microscopic calculations of the electronic surface structure in the presence
of a magnetization domain wall for our model TCI.  Our goal is to explicitly demonstrate the
presence of gapless, chiral conducting states within the surface energy gap
generated by a uniform magnetization, as found in the previous section.  We will
see that the number for each chirality agrees with our expectations based on the
Chern number calculations, and see that these are robust for different microscopic
realizations of the DW magnetization profiles.  The numerical approach will also
allow us to assess the energy of a DW excitation, which is of particular interest
in the context of situations where the ground state magnetization is along a
$\Gamma$-$L_i$ direction, with $i=1,2,$ or 3.  These directions are associated with
the $\bar{M}$ points in the surface BZ, and there are six degenerate groundstate directions
when the chemical potential is adjusted near the energy of the Dirac points associated
with these locations \cite{Reja_2017}.  These directions however come in two groups of
3, with components of the magnetization perpendicular to the surface either directed upward
or downward.  A priori it is unclear whether DW's connecting states with the same
perpendicular component or opposite ones is lower in energy; in our model we will see
that the latter is lower in energy.  This means that the system in these circumstances
should be regarded as a six state system, rather than one with two sets of three states with
a relatively large barrier separating states in different groups.

We begin by explaining how the numerical calculations are carried out.

\subsubsection{Projected Hamiltonian in Presence of Domain Wall}

Our basic approach is to create a Hamiltonian with magnetization on the surfaces varying
with position, to form a DW configuration.  This means we will be working with very large
unit cells, so that computation of the electron states becomes impractical for the full
set of states in the slab geometry.  We thus continue to exploit the technique of
projecting the Hamiltonian into the low energy space of surface states.
For simplicity we consider DW's which run along the two highest symmetry directions
on the surface, along the $k_1$ and $k_2$ directions illustrated in Fig. \ref{fig_bz}(c).
Our supercells are very large along the cross-sectional direction of the DW,
but as the magnetization is a function of displacement in only one direction,
they can be very small in the direction perpendicular to this.
Because the real space atoms on the surface are laid out in a triangular lattice,
neighboring atoms in general will have displacements both parallel and perpendicular
to the DW cross-section.  To deal with this we allow our supercells to have a width
containing two atoms along the narrow direction [see Fig. \ref{fig_bz}(b)], so that the magnetization need depend
only on the position of an atom along the cross-sectional direction.

Thus, the supercell will be constructed of a line of small unit cells, defined by the primitive
lattice vectors $a_1$ and $a_2$ shown in Fig. \ref{fig_bz}(b).  The BZ associated with this
doubled unit cell can be represented by a rectangle, as shown in (c) of the same figure.
Notice this is half the size of a unit cell containing only one surface atom, so that
$\bar{M}$ points of the latter falling outside of the former get folded in.
In particular this means the $\bar{M}_1$ point
will coincide in the smaller BZ with the $\bar{\Gamma}$ point, and the $\bar{M}_2$ and $\bar{M}_3$
points will coincide with one another.

We next need to generate a set of basis states that can represent a magnetization profile that
varies slowly over many 2-atom unit cells.
As a concrete example, suppose that the
magnetization rotates as we move along the $a_1$ direction.
If we impose periodic boundary conditions, we are required to have {\it two}
DW's separating regions of uniform magnetization in different directions. Let
$N_c$ be the number of unit cells within which the full profile is contained.
Our basis is generated for this large supercell in the absence any magnetic
moments, by fixing $k_2$, and diagonalizing the Hamiltonian
for a unit cell of the slab with only 2 surface sites, and with
quantized values of $k_1$ of the form
$k_1=k_m=2\pi m/N_c; m=0,1,2,...(N_c-1)$.
For each momentum, we retain only $N_s$ states with energies closest to the bulk gap,
which capture the surface states.  (Typically $N_s=8$ works well in our calculations.)
We thus retain $N_c$x$N_s$ states in total for each of the quantized $k_1$ momenta.
These basis states may be represented as
\begin{eqnarray}
|k_m,j\rangle &\equiv & \sum_{i}\alpha_{k_m}^j(i)|\alpha_i\rangle \nonumber\\
&=&\frac{1}{\sqrt{N_c}}\sum_{n}e^{-ik_m x_n}
|n,j\rangle\
\end{eqnarray}
where $|\alpha_i\rangle\equiv |i_s,o_i,s_i\rangle$ represents basis states indexed by site $i_s=1,2,...2$x$47$,
$o_i=(p_x, p_y, p_x)$ the orbital index, and $s_i=(\uparrow,\downarrow)$ the local
spin index. The quantities $x_n$ denote the positions of the two atom unit cells within the larger supercell.
Thus
for each $k_m$, we have retained $j=1,2,...N_s$ states, which
we will use for the basis of our Hilbert space. The energy eigenvalue (again, in the absence of
any magnetization) for the state $|k_m,j\rangle$ is denoted by
$E_{k_m,j}$.

Rewriting our basis in real space by inverting the Fourier transform,
\begin{eqnarray}
|n,j\rangle=\frac{1}{\sqrt{N_c}}\sum_{k_m,j}e^{ik_m x_n}
|k_m,j\rangle,\
\end{eqnarray}
we can now introduce surface magnetic moments into the Hamiltonian, writing
as $H_n$ the projection of the $sd$ Hamiltonian for
the two sites in the cell located at $x_n$, with each site containing
the values of ${\bf S}_i$ determined by the presumed magnetization profile of the DW.
With this addition, the effective Hamiltonian matrix for our system becomes
\begin{widetext}
\begin{eqnarray}
\langle k_m,j | H_{eff} | k_{m'},j'\rangle
=\frac{1}{N_c}\sum_{n=0}^{N_c-1}\langle k_m,j | H_{n} |
k_{m'},j'\rangle
e^{i{(k_m-k_{m'})}x_n}+E_{k_m,j}\delta_{k_m,k_{m'}}\delta_{j,j'}.
\label{eq_proj_H_mag}
\end{eqnarray}
\end{widetext}
Note again that this matrix is dependent implicitly on the value of $k_2$, the
wavevector in the direction along which the DW runs.
This matrix is considerably reduced in size from what one has for the tight-binding model
of the full slab with a magnetization profile on its surface, and allows us to compute
energy states of the electrons as a function of $k_2$.  For DW's running along the
$a_1$ direction,
we construct an effective Hamiltonian in a very analogous way.


\begin{figure}
  \centering
    \includegraphics[width=0.99\linewidth]{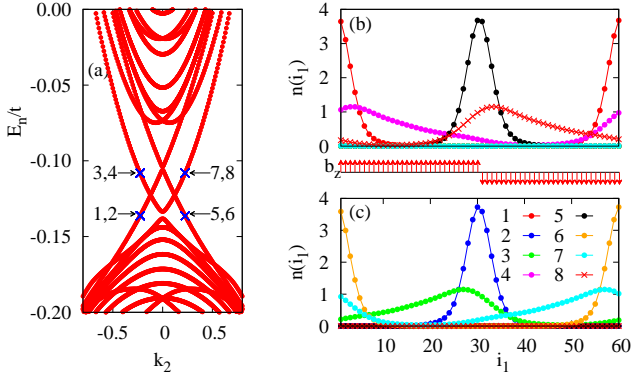}
    \caption{(a) Energy bands $E_n$ near $\bar{\Gamma}$ when a Neel domain wall of width $d=0$ runs along $k_2$. The electron density $n(i_1)$ for the representative DW states (indicated by blue cross points and numbered $1,2...$) along the unit cell direction $a_1$ are shown in (b) and (c) for top and bottom surfaces respectively of the slab with (111) surfaces. The component of magnetic moments $b_z$ along \protect\glzero direction are shown by red arrows between panels (b) and (c)
    .}
    \label{fig_gbar_d_0_dw_paralell_k2_Gbar}
\end{figure}

\begin{figure}
  \centering
    \includegraphics[width=0.99\linewidth]{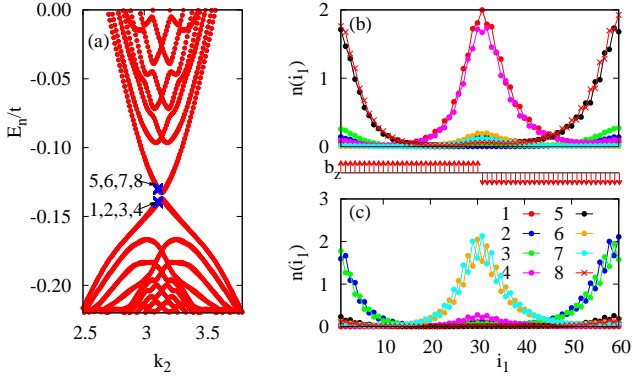}
    \caption{Energy bands $E_n$ [panel (a)]  and electron densities [panels (b), (c)] of bound DW
states near the $\bar{Y}$ point for the
DW configuration as described in Fig. \ref{fig_gbar_d_0_dw_paralell_k2_Gbar}.}
    \label{fig_gbar_d_0_dw_paralell_k2_Ybar}
\end{figure}

\subsubsection{Results}
With this formalism, we now compute electronic structures for
different DW configurations. We expect to find states invading the
gaps present in the surface electronic structure when there is
a uniform magnetization. These occur near two places
(see Fig. \ref{fig_bz}).
(i) The center of the rectangular BZ where $\bar{\Gamma}$ and $\bar{M}_1$ overlap due to zone-folding.
(ii) The projection of the $\bar{M}_2$ and $\bar{M}_3$ points onto the $k$-axis
running along the DW. The latter corresponds
to either the $\bar{X}$ or the $\bar{Y}$ point in the
reduced Brillouin zone shown in Fig. \ref{fig_bz}(c), depending on which direction
the DW runs along.
We will see that the in-gap
states appear when the projection of the magnetic moments along
any of the bulk $\Gamma$-$L$ directions changes sign inside the DW cross-section.
We expect from our Chern number analysis that
the number of in-gap branches depends on the number of such projections changing sign.

We first consider the case of
DWs connecting different states with magnetic moments along the $\Gamma$-$L_0$ axis,
with the DW's running along the $a_2$ direction.
[See Fig. \ref{fig_bz}(b)].
In this case the magnetic moments rotate as we move in the $a_1$ direction within a DW,
and the rotation is in the plane defined by the direction perpendicular
to the surface and the $a_1$ direction.  This represents a N\'eel domain wall \cite{Chaikin_1995}.
The geometry of our supercell includes two regions of width $N_s-2d$ with uniform magnetization,
one pointing ``up'' and the other ``down'', connected by two DW's of width $d$ within which the magnetization
rotates uniformly.  We consider several values of $d$, including $d=0$ for which the
change in magnetization is abrupt.


Fig. \ref{fig_gbar_d_0_dw_paralell_k2_Gbar}(a) illustrates the band structure near the $\bar{\Gamma}$ point
as a function of $k_2$ for a DW with $d=0$ and $|J{\bf S}|=0.1$. As noted above,
the $\bar{\Gamma}$ point hosts two Dirac points, associated with the surface
projections of the bulk $L_0$ and $L_1$ points,
due to zone-folding of the original hexagonal Brillouin zone
[Fig. \ref{fig_bz}(c)]. Since this DW configuration induces a sign change in the component of magnetic moments
along the \glzero and the \glone directions, we expect to find two chiral states associated with these.
Because we have two surfaces, each with two DW's, this leads to an expectation of 8 chiral states.
Fig. \ref{fig_gbar_d_0_dw_paralell_k2_Gbar}(a) shows this is indeed true.  (Note each of the states
in the figure is exactly doubly degenerate, due to a combination of time-reversal and inversion symmetries.)
Figs. \ref{fig_gbar_d_0_dw_paralell_k2_Gbar}(b) and (c) show the electron densities of representative states
from the different chiral branches,
for each of the DW's on the top and bottom surface.  It is clear that each of the DW's
hosts two chiral states, running in opposite directions.  This is consistent with the Chern
number change we found in the last section, which was $\pm 1$ for the $\bar{\Gamma}$ point,
and $\mp 1$ for a $\bar{M}$ point.  Note the small gap opening at $k_2=0$ near energy -0.14 occurs
due to admixture of DW states associated with the $\bar{M}$ point on the same surface:
as the densities in Figs. \ref{fig_gbar_d_0_dw_paralell_k2_Gbar}(b,c) show, the
localization lengths for these states are still relatively large compared to our
inter-DW separation, even for the large unit cells we use.  This is a reflection of
the fact that within the uniformly magnetized regions, the magnetization is not parallel
to the \glone direction, so the gaps induced in the Dirac points at $\bar{M}$ are relatively small.


In contrast, the band structure near $\bar{Y}$
associated with this magnetization profile
yields states in each DW with the {\it same} chirality.  This is
shown in Fig.\ref{fig_gbar_d_0_dw_paralell_k2_Ybar}.  For example, the states labeled 1 and 4
disperse in the same direction, and are located in the {\it same} DW.
Analogous calculations (not shown) of DW's
running perpendicular to the structure relevant for Figs.
\ref{fig_gbar_d_0_dw_paralell_k2_Gbar} and \ref{fig_gbar_d_0_dw_paralell_k2_Ybar} yield
analogous results.
We thus confirm that the net chirality of DW states connecting groundstates with
magnetizations along the \glzero axis, but in opposite directions, have net
chirality of 2.  This is just as expected from our Chern number analysis.

\begin{figure}
  \centering
    \includegraphics[width=0.99\linewidth]{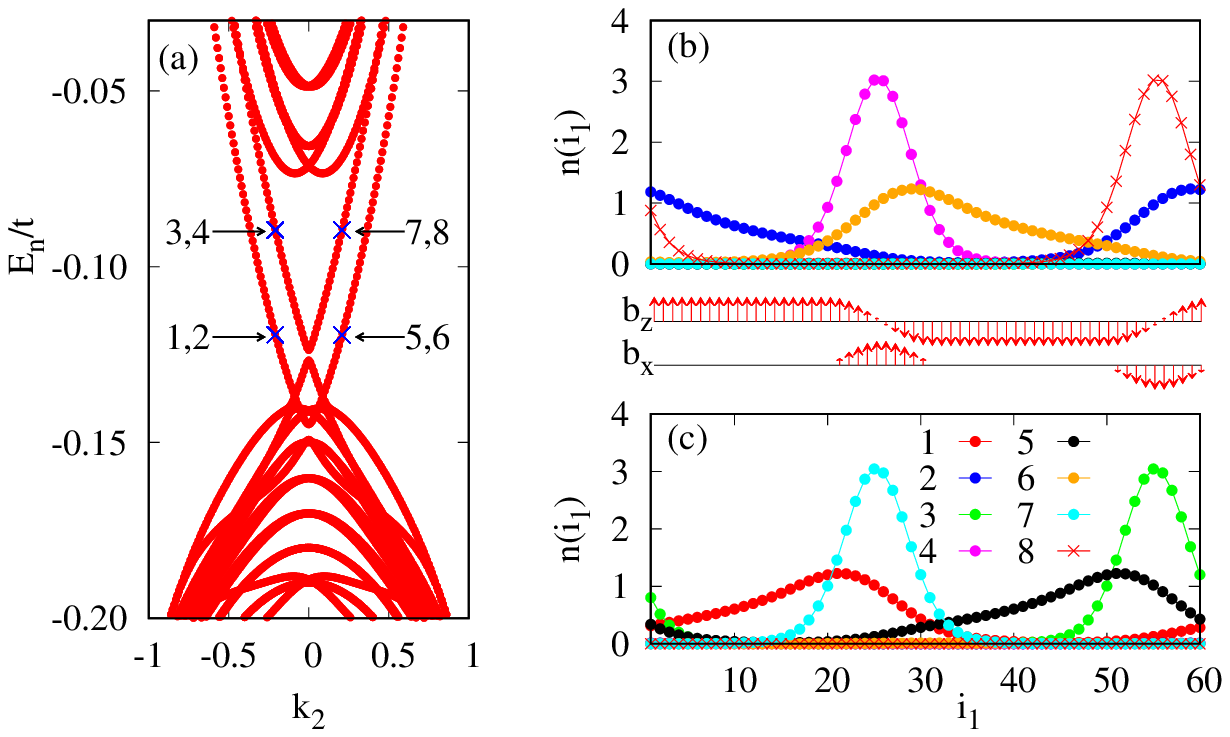}
    \caption{(a) Energy bands $E_n$ near $\bar{\Gamma}$ when a Neel domain wall of width $d=10$ runs along $k_2$. The electron density $n(i_1)$ for DW states (indicated by blue cross points and numbered $1,2...$) along the unit cell direction $a_1$ are shown in (b) and (c) for top and bottom surfaces respectively of the slab with (111) surfaces. The component of magnetic moments $b_z$ along \protect\glzero and $b_x$ along $a_1$ directions are shown by red arrows between panels (b) and (c).}
    \label{fig_gbar_d_10_dw_paralell_k2_Gbar_Neel}
\end{figure}

\begin{figure}
  \centering
    \includegraphics[width=0.99\linewidth]{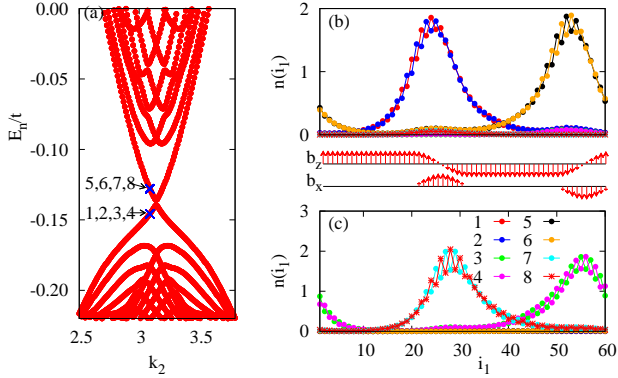}
    \caption{
Energy bands $E_n$ [panel (a)]  and electron densities [panels (b), (c)] of bound DW
states near the $\bar{Y}$ point for the
DW configuration as described in Fig.\ref{fig_gbar_d_10_dw_paralell_k2_Gbar_Neel}.
}
    \label{fig_gbar_d_10_dw_paralell_k2_Ybar_Neel}
\end{figure}

Further analogous calculations may be carried through for other geometries.  For example,
Figs. \ref{fig_gbar_d_10_dw_paralell_k2_Gbar_Neel} and \ref{fig_gbar_d_10_dw_paralell_k2_Ybar_Neel}
illustrate results for wider DW's, $d=10$.  The results are qualitatively very similar to our
$d=0$ results, importantly showing the same types of chiral states near the $\bar{\Gamma}$
and $\bar{Y}$ points as for $d=0$, and the same net chirality for the DW's that we expect
based on the Chern number analysis.  We have found other values of $d$, both larger and smaller,
yield these types of results as well.  In addition we have performed calculations for Bloch walls --
profiles in which the rotation axis of the magnetization inside the DW is parallel rather than
perpendicular to the direction along which the DW runs -- and again find the same basic results.
As might be expected for topologically determined properties, the chirality of DW's in
this system seems rather robust.

\begin{figure}
  \centering
  \includegraphics[width=0.99\linewidth]{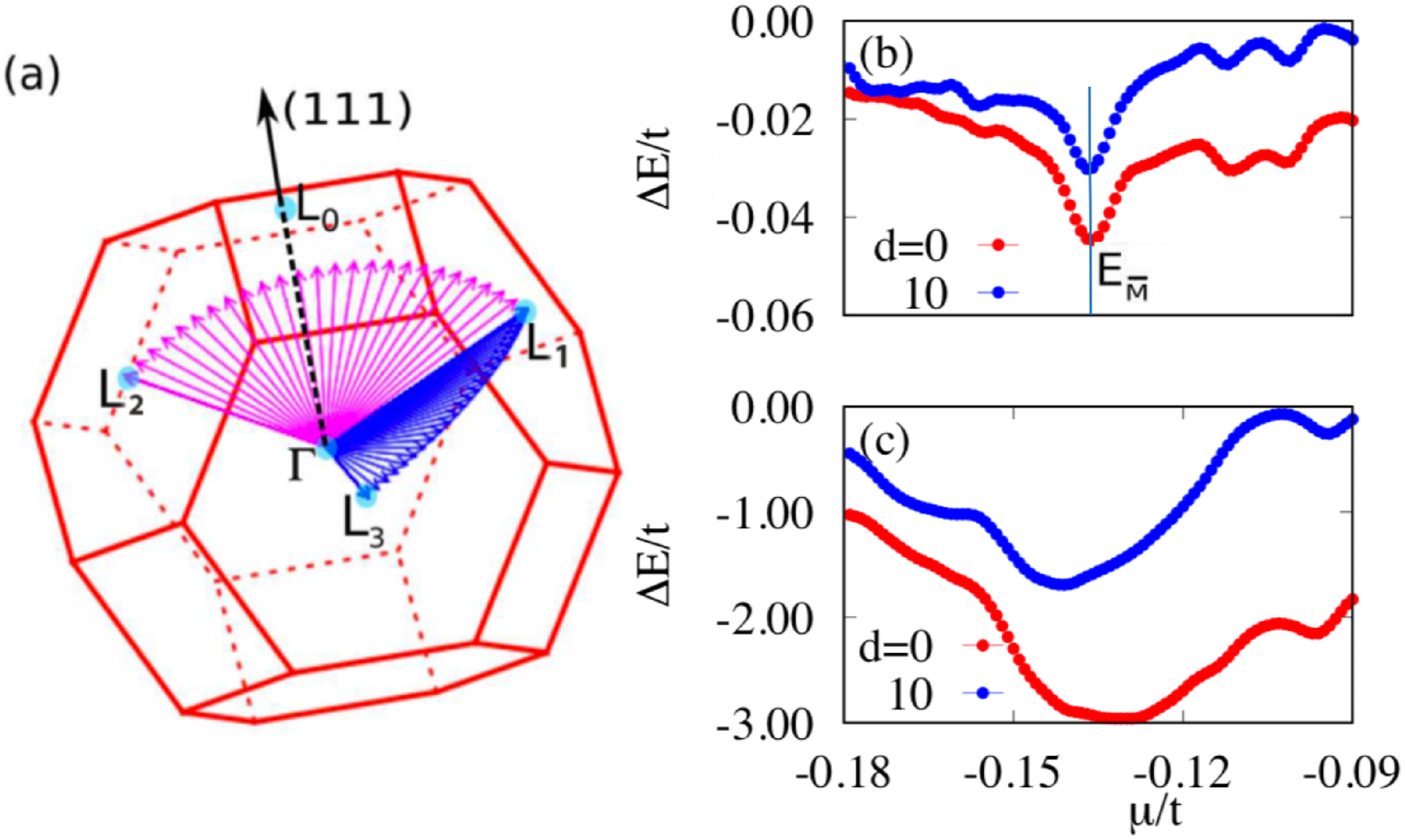}
    \caption{(a) DW configurations connecting (i) $\Gamma-L_1$ to $\Gamma-L_2$ directions (magenta), (ii) $\Gamma-L_1$ to $\Gamma-L_3$ directions (blue). (b) The energy difference between these configurations $\Delta E=E_{ii}-E_i$ as a function of chemical potential $\mu$ for $d=0,10$ and $|J{\bf S}|=0.01$ showing minimum when $\mu$ is close to
$\bar{M}$ Dirac point energy $E_{\bar{M}}$. (c) $\Delta E$ for larger $|J{\bf S}|=0.10$}
\label{fig_energy_diff}
\end{figure}

We also wish to consider DW's connecting different states associated with magnetization groundstates
along the $\Gamma-L_{1,2,3}$.  These are energetically stable when the chemical potential is near
the energy of the Dirac points associated with $\bar{M}$ points.  As mentioned above, what
is not a-priori obvious is whether DW's that connect groundstates with the same sign of
of magnetization along the direction perpendicular to the surface will be higher or lower
in energy than those connecting neighboring magnetization states with opposite such projections.
Our calculations support that it is in fact the second of these that is energetically favorable.
To show this, we
consider DW configurations as shown in Fig. \ref{fig_energy_diff}(a).
There are two cases: (1) one which  connects
the \glone to \gltwo directions (magenta), and (ii) one which connects the
\glone direction to the \glthree direction (blue).  Using the technique described above, we compute the
single-particle energy states for each of the two structures, and then add all the energies below
the Fermi energy $\mu$ to obtain a total energy associated with the magnetization profile.
The energy difference of these,
$\Delta E=E_{ii}-E_i$, as a function of $\mu$,
is shown in Fig.\ref{fig_energy_diff} (b) for $|J{\bf S}|=0.01$ and $d=0, 10$ as indicated.
We find that the DW configuration (ii) is favorable over (i), and moreover that $\Delta E$ has a local
minimum, when $\mu$ is close to the $\bar{M}$ Dirac point energy, $E_{\bar{M}}$.
This has the important consequence of making all six groundstate configurations equally accessible
from some given starting state, yielding a six state clock model.  If $\mu$ is near $E_{\bar{M}}$
we expect, as discussed above, that system will thermally disorder
at sufficiently high temperature via a Kosterlitz-Thouless transition \cite{Jose_1977}.

\begin{figure}
  \centering
    \includegraphics[width=0.99\linewidth]{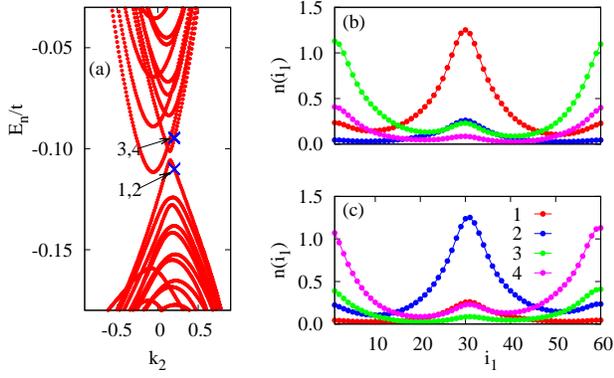}
    \caption{(a) Energy bands $E_n$ at $\bar{\Gamma}$ for DW configuration 
    corresponding to ($ii$) in Fig. \ref{fig_energy_diff}(a), with $d=0$. The density of electron $n(i_1)$ for the representative DW states (indicated by blue cross points and numbered $1,2...$) along the unit cell direction $a_1$ are shown in (b) and (c) for top and bottom surfaces respectively of the slab with (111) surfaces. }
    \label{fig_mbar_d_0_dw_paralell_k2_Gbar}
\end{figure}

\begin{figure}
  \centering
    \includegraphics[width=0.99\linewidth]{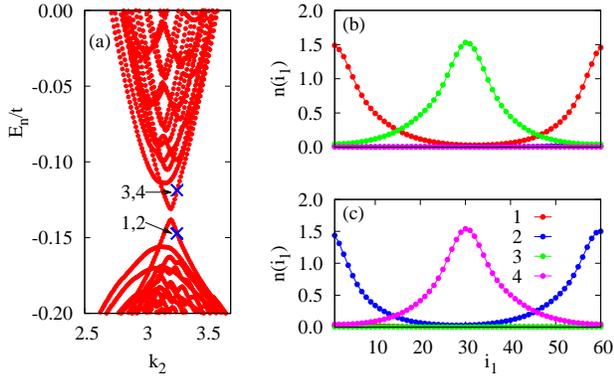}
    \caption{(a) Energy bands $E_n$ near $\bar{Y}$ for DW configuration
    corresponding to ($ii$) in Fig. \ref{fig_energy_diff}(a), with $d=0$, for momentum along $k_2$. The electron density $n(i_1)$ for the representative DW states (indicated by blue cross points and numbered $1,2...$) along the unit cell direction $a_1$ are shown in (b) and (c) for top and bottom surfaces respectively of the slab with (111) surfaces. }
    \label{fig_mbar_d_0_dw_paralell_k2_Ybar}
\end{figure}

Finally, it is interesting to contrast the nature of the in-gap states hosted by these DW's with those
relevant to magnetizations along the \glzero axis.  Examination of Fig. \ref{fig_energy_diff}(a)
reveals that while the magnetization projection along the \glone and \glthree directions does
not change sign, those along the \glzero and \gltwo directions do.  This means for a DW running
along the $k_2$ direction, we should find a single chiral state near each of the $\bar{\Gamma}$
and $\bar{Y}$ points.  Figs. \ref{fig_mbar_d_0_dw_paralell_k2_Gbar} 
and \ref{fig_mbar_d_0_dw_paralell_k2_Ybar} demonstrate that this
indeed happens.  Note that the chiral directions of the two modes are {\it oppositely} oriented
within a given DW, so that the net chirality vanishes.  This is consistent with our observation,
in the previous section,
that the $\bar{M}$ and $\bar{\Gamma}$ Chern numbers have opposite signs.  This can have interesting
consequences for differing electrical behaviors due to DW's when $\mu$
is near the energy of the Dirac points at $\bar{M}$ as opposed to that of the $\bar{\Gamma}$ point.
We discuss this further in the next and final section of this paper.



\section{Summary, Discussion, and Future Directions}

In this paper we studied domain walls of ferromagnetic systems, in which the magnetic
degrees of freedom mutually interact through their impact on Dirac electrons on a surface.
Such models arise naturally in the context of topological
insulators protected by time-reversal symmetry (TI's) and topological crystalline insulators (TCI's),
and are very
commonly studied perturbatively, using varieties of the RKKY analysis.  In our study we
demonstrated that if magnetic order does set in this type of system, the energetics of magnetization gradients
may become anomalous, in a way that is in principle controllable.  When the surface electron
density is such that there is a Fermi surface, the interactions effectively cut off at
a length scale of order $1/k_F$, above which there are $2k_F$ oscillations in the RKKY coupling.
As $k_F \rightarrow 0$, the coupling retains its sign,
and the RKKY analysis predicts a (well-known) $1/R^3$ fall-off in the coupling.
In a coarse-grained description of the system, this means that the appropriate
gradient term for the magnetization at low temperature
becomes anomalous, acquiring an emergent long-range form, with true long-range
interactions among magnetization gradients being the limiting behavior as the magnetization
magnitude vanishes. For non-vanishing scale of magnetization, the gradient energy can be
properly described by a form that is quadratic in wavevector, but acquires a non-analytic form in the
magnetization itself.

The emergent long-range form of the interaction impacts, among other things, interactions among
DW's, since these involve a fixed change in magnetization.  From our analysis of the gradient
energies, we showed that the emergent interaction induces logarithmic interactions between
DW's, up to a length scale set by the magnetization itself.  Using an effective Dirac model
in conjunction with a transfer matrix method, we were able to verify the presence of this
interaction, and found moreover that it results from a subtle cancellation in the energies
associated with bound states in the DW's and phase shifts of unbound electrons scattering
from them.  A tight-binding system involving graphene with a position-dependent mass term
that models DW pairs corroborated the result.

We then considered DW's in a more concrete system, a model of (Sn/Pb)Te alloys that are
a paradigm for TCI systems.  We considered the (111) surface, which hosts particularly rich
physics in this context, because it hosts Dirac points at two different, distinct energies,
a single isolated Dirac point (near the surface $\bar{\Gamma}$ point) and, at slightly lower energy,
a group of three degenerate
Dirac points (near three $\bar{M}$ points), allowing for different types of DW's.
We carried out a numerical Berry's phase analysis
on the electronic states around these points in the presence of a uniform magnetization,
and demonstrated that they carry Chern numbers of opposite sign, $\pm1/2$.  When the chemical
potential is adjusted such that the $\bar{\Gamma}$ point dominates the energetics of
the magnetization, the resulting DW excitations are predicted to induce a change of
Chern number given by $\pm 2$.  This suggests the DW's host in-gap states with a net
chirality.  We demonstrated that this is true using a numerical low-energy projection scheme for
the tight-binding slab, and showed that it arises as a net effect of {\it four} in-gap
states, with two running in opposite directions, and another pair running in the same
direction.  For cases where the $\bar{M}$ Dirac points dominate the magnetization energy,
we found that the lowest energy DW's of equally connect six possible groundstate orientations,
and in this case yield {two} conducting states of opposite chirality.

\begin{figure}
  \centering
    \includegraphics[width=0.99\linewidth,trim = 250 180 150 200,clip]{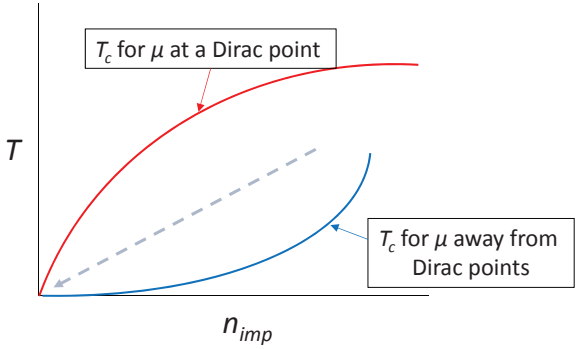}
    \caption{Schematic phase diagram for classical magnetic impurities coupled
by surface Dirac electrons, contrasting
behavior of the critical temperature $T_c$ vs. impurity density $n_{imp}$ when the
Fermi energy $\mu$ passes through a Dirac point (red) with when it does not (blue).
Dashed illustrates illustrates a trajectory in which the in-gap states associated with domain
walls will produce pseudogap behavior in the electronic spectrum, and where the
interactions among the DW's become increasingly long range moving down the trajectory,
enhancing the density of states within the mean-field gap.}
    \label{Tcfig}
\end{figure}

The conducting states of DW's in these systems are of considerable interest, because they
allow their presence to be detected electrically.  DW's can be forced into the system, for example,
by cooling it from high temperature in zero field.  The DW's can be detected in principle by a variety of
techniques, by looking for their contribution to the conductance of the surface.  This could
be investigated by transport studies, tunneling measurements, or even surface reflectance.
The behavior of the system as the chemical potential is changed should reveal the different
regimes of the low-energy DW's, as the system is tuned through different
behaviors of the gradient energy, as well as through Fermi energy scales where different Dirac
points may dominate the magnetization dynamics.  It is interesting to note, for example,
that in the two-fold case (one low-energy magnetization axis) the DW's should be strongly
conducting due to their chirality, whereas in the six-fold case, the vanishing chirality
will allow backscattering between the in-gap states in a DW, leading to a smaller
contribution to the surface conductance.  Beyond this,
an estimation of the critical temperature $T_c$, based on balancing of energy and entropy
of a DW,  reveals a crossover from a $T_c \sim n_{imp}^2$  when the
Fermi surfaces are closed loops to $T_c \sim n_{imp}$ when there is a point-like Fermi
surface (i.e., when the Fermi energy passes through a Dirac point.)  This behavior
is illustrated in Fig. \ref{Tcfig}.

While our detailed analyses of these systems have largely focused on the low-energy behavior of
the topological DW excitations, it is interesting to consider the consequences of our results for
higher temperatures.  In particular, approaching a phase boundary for some magnetic impurity
density $n_{imp}$, $T_c(n_{imp})$, one expects the average magnetization to become vanishing small
as DW's come increasingly close to proliferating.
However, this does not imply that the interactions among the DW's become unlimited in range:
in such a situation, the stiffness becomes limited by $k_BT$, rather than the magnetization
scale \cite{note2}.  For example, a calculation akin to that of Sec. \ref{sec:graphene_stiffness} at
finite temperature $T$ reveals that the energy cost to introduce an magnetization
gradient $g$ in graphene behaves as $\Delta E(g) \sim g^2b_0^2/T$ \cite{unpub}.  This indicates that
the long-range behavior of the stiffness will be cut off by finite temperature if
the magnetization scale is small.  Thus, true long-range interactions in this system emerge
if one approaches the low-temperature, low impurity density point, as illustrated in by
the dashed arrow in Fig. \ref{Tcfig}.  In approaching this point, interactions among DW's
of unlimited range emerge.  It is interesting to note that the in-gap states hosted
by finite size DW's will in principle fill the mean-field gap in the Dirac electron spectrum,
but the density of states associated with these will drop rapidly approaching zero energy
as DW's of increasing size (which will host the lowest energy in-gap states) are exponentially
unlikely to be found in the system when in the ordered state.  The emergent long-range interactions
will enhance the average area occupied by DW's and their associated induced states in the gap relative to systems
with short-range gradient interactions.  In principle, this behavior should be directly accessible
in tunneling experiment.

The studies we have reported in this paper suggest many other directions for future exploration.
Important among these is that,
in our approach to these systems, we have treated the magnetic moments as classical.
Clearly at sufficiently
low temperature a quantum treatment would be more appropriate.
For example, we have ignored the possibility of non-trivial correlations
between conduction electron spins and the impurity spins that occur
in the Kondo effect, although this physics should set in at extremely
low temperature when the $sd$ coupling scale $J$ is small \cite{Allerdt_2017}.
Beyond this, it is interesting
to note the connection of this system with ``chiral magnets,'' \cite{note1} magnetic systems coupled
to chiral fermions \cite{Rosenstein_1993,Knorr_2016,Otsuka_2016,Li_2017,He_2018,Yin_2018}, which are known
to support quantum phase transitions with their own unique critical behaviors.  Note that while
such systems are similar to the ones we focus upon, these are generally formulated
as magnets supporting their own independent gradient interactions, exchange-coupled
to chiral fermions, while in the systems we are considering, interactions among the
magnetic moments arise {\it solely} from exchange coupling with the Dirac electrons.
From the perspective of an renormalization group (RG) analysis the
systems may be connected, in which case the origin in Fig. \ref{Tcfig} will move
to a non-vanishing value of $n_{imp}$.  The classical behavior discussed in
our work will nevertheless present itself as crossover behavior prior to quantum critical behavior
sufficiently close to the transition point.  Our studies demonstrate that interesting
fluctuation behavior appears in this system even away from the quantum critical regime.

Related to this, systems such as graphene, in which spin-orbit coupling is largely
irrelevant so that the magnetization enjoys continuous symmetries, offer further
possibilities for study.  Interacting electrons in graphene without external
magnetic moments can be formally recast in terms of non-interacting electrons with
an auxiliary Hubbard-Stratanovich field \cite{Knorr_2016}, suggesting a quantum phase transition
in the universality class of the Gross-Neveu model \cite{Herbut_2006}.  How this picture
changes when real quantum spins couple to the electrons remains an interesting
area to investigate.
While the continuous symmetry of the order parameter implies that
thermal fluctuations at any non-zero temperature disorder the system \cite{Mermin_1966},
the non-analytic behavior of the system with respect to spin gradients at
short wavelengths suggest that interesting collective modes can be present
in this regime.  Moreover, the effect of thermal disordered magnetic moments
on the electron states of this system should have interesting consequences for
thermal and transport properties of the system.


Finally, effects of disorder have been assumed throughout this paper to sufficiently
average that its effects may be ignored at a qualitative level.
This seems most likely for situations where the effective interactions have become sufficiently
long range,
but when the interaction length scale is fixed by a Fermi momentum,
they are likely to become more important.  In addition, electron-electron interactions
have been ignored throughout our study.  In systems where Fermi surfaces and Dirac points
may coexist at the same energy -- such as the (111) TCI surface -- these will be screened and are
likely to be qualitatively unimportant.  Other surfaces, such as TI systems or the (100) surface
of the (Pb/Sn)Te TCI system, can become fully gapped, and here we expect logarithmic,
repulsive interactions among DW's because of the charge they contain. These interactions
will be present to arbitrarily large distance even at finite $T$, and whether they
impact classical {\it thermal} phase transitions in these systems
is another interesting direction to explore.

Clearly,  magnetic degrees of freedom coupled by Dirac electrons host a rich
variety of physical phenomena.  Under many circumstances, these systems
support domain walls as fundamental topological excitations, which reflect the interesting effective
interactions induced among the magnetic moments, as well as the topological nature of
the electronic system that couples them.  Their behavior, both thermal and
electrical, offer exciting windows into the special properties of
electrons in such topologically non-trivial systems.

\textit{Acknowledgements --} The authors gratefully acknowledge useful discussions with Ganpathy Murthy,
R. Shankar, Efrat Shimshoni, Kai Sun, and Shixiong Zhang.
This work was supported by the NSF through Grant Nos.
DMR-1506263 and DMR-1506460, by the US-Israel Binational Science Foundation, and by
MEyC-Spain under grant FIS2015-64654-P.  HAF thanks the Aspen Center for Physics, where
part of this work was done.  Computations were carried out on the
ITF/IFW and IU Karst clusters.

\begin{widetext}

\section{Appendix}

In this Appendix we provide a few details of the stiffness calculations whose results are described
in Section \ref{sec:gradients}.  We begin first with the case where the Fermi energy is in the
gap, from Eq. \ref{DeltaEQ}, which we reproduce for convenience:

\begin{equation}
\Delta E = - {1 \over 4} \sum_{\bf q} \left\{
\frac{|\langle {\bf q},- | \delta {\bf b} \cdot \vec{\sigma}|{\bf q}-{\bf Q},+\rangle|^2}{\varepsilon_0({\bf q}) +
\varepsilon_0({\bf q}-{\bf Q})} +
\frac{|\langle {\bf q},- | \delta {\bf b} \cdot \vec{\sigma} |{\bf q}+{\bf Q},+\rangle|^2}{\varepsilon_0({\bf q}) + \varepsilon_0({\bf q}+{\bf Q})} \right\}.
\end{equation}
To find the gradient energy
we expand this to quadratic order in $Q$.  A long but in principle straightforward calculation
brings us to the expression
\begin{eqnarray}
\Delta E({\bf Q}) - \Delta E(0) \approx
{1 \over {32}}\sum_{\mu,\nu=x,y} Q_{\mu}Q_{\nu}\sum_{\bf q} \Biggl\{
\frac{|\langle {\bf q},- | \delta {\bf b} \cdot \vec{\sigma}|{\bf q},+\rangle|^2}{\varepsilon_0({\bf q})^2}
\partial_{\mu}\partial_{\nu} \varepsilon_0({\bf q}) \nonumber\\
-{1 \over {\varepsilon_0({\bf q})}} \partial_{\mu}\partial_{\nu}
|\langle {\bf q},- | \delta {\bf b} \cdot \vec{\sigma}|{\bf q},+\rangle|^2
\Biggr\rbrace
\label{DeltaEQsmall}
\end{eqnarray}
This expression is explicitly quadratic in $Q$ {\it and}
$\delta b$.
As discussed in the main text
it is natural to introduce a tensor $g_{\mu\nu}^{ij}$
characterizing the energy cost,
so that $\Delta E({\bf Q}) - \Delta E(0) =
{{\Omega} \over {2}}\sum_{\mu,\nu=x,y} \sum_{ij=x,y,z} g_{\mu\nu}^{ij}
Q_{\mu}Q_{\nu}\delta b_i \delta b_j$.  The $g$
coefficients can read off from Eq. \ref{DeltaEQsmall},
and for fixed $\delta \bf{b}$ one can use them to assess the
energy cost for introducing a slow gradient in the magnetization.
More
explicit expressions for the $g$'s require
a matrix element, which can found using Eq. \ref{spinor}.  This yields
\begin{equation}
|\langle {\bf q},- | \delta {\bf b} \cdot \vec{\sigma}|{\bf q},+\rangle|^2 =
\left\lbrace \left[q^2\delta b_z - b_z {\bf q} \cdot \delta {\bf b} \right]^2 +
\varepsilon_0(q)^2\left[\hat{z}\cdot({\bf q} \times \delta {\bf b}) \right]^2
\right\rbrace /[ q\varepsilon_0(q)]^2,
\label{squareME}
\end{equation}
which in turn provides integral expressions of the form
\begin{eqnarray}
\sum_{ij} g_{\mu\nu}^{ij}\delta b_i \delta b_j &=&
{{1} \over {\Omega}}\sum_{\bf q}
\frac{2\delta_{\mu\nu}-4q_{\mu}q_{\nu}/\varepsilon_0(q)^2}
{q^2\varepsilon_0(q)^5} \nonumber \\
&\times&
\left\lbrace
q^4\delta b_z^2 + b_z^2 \left( q_x^2\delta b_x^2 + q_y^2 \delta b_y^2 \right)
+\varepsilon_0(q)^2 \left( q_x^2\delta b_y^2 + q_y^2 \delta b_x^2 \right)
-2q^2q_xq_y\delta b_x \delta b_y \right\rbrace.
\label{rho_integrals}
\end{eqnarray}
It is immediately apparent that only $g_{\mu\nu}^{zz}$ and
$g_{\mu\nu}^{ij}$ with $i,j=x,y$ are non-vanishing, so that gradients
in $\delta b_z$ can be assessed separately from gradients in $\delta b_{x,y}$.
The various non-vanishing values of $g_{\mu\nu}^{ij}$ can now be
read off in integral forms, all of which are analytically tractable.
The explicit results are given in Eqs. \ref{g33} and \ref{g_in_plane}.

We next consider the case when the Fermi energy passes through a band.  Our starting
point is now the expressions for $\Delta E_+$ and $\Delta E_-$.  The former is given
by Eq. \ref{DeltaEpl}, which again we reproduce for convenience:
\begin{equation}
\Delta E_+ =
{1 \over 4} \sum_{\mathclap{\substack{q>k_F \\
|{\bf q} -{\bf Q}|<k_F}}}
\frac{|\langle {\bf q},- | \delta {\bf b} \cdot \vec{\sigma}
|{\bf q}+{\bf Q},-\rangle|^2}{\varepsilon_0({\bf q}+{\bf Q}) - \varepsilon_0({\bf q})}
-{1 \over 4} \sum_{q>k_F}
\frac{|\langle {\bf q},- | \delta {\bf b} \cdot \vec{\sigma}
|{\bf q}+{\bf Q},+\rangle|^2}{\varepsilon_0({\bf q}+{\bf Q}) + \varepsilon_0({\bf q})},
\end{equation}
and again
$\Delta E_-$ has the same form as Eq. \ref{DeltaEpl}, with ${\bf Q} \rightarrow -{\bf Q}$.
The constraints on the wavevector sums can be simplified by defining a
step function,
\begin{equation}
f_{\bf q}=
\Biggl\{
\begin{array}{c c}
0 & \quad q < k_F, \\
1 & \quad q > k_F,
\end{array}
\end{equation}
and a unit vector
$\hat{h}_{\bf q} \equiv (q_x,q_y,b_z)/\varepsilon_0(q)$.
Using Eq. \ref{spinor} to compute the matrix elements,
with considerable algebra one can reformulate $\Delta E$ as
$\Delta E \equiv \Delta E_+^{(1)}+\Delta E_+^{(2)}
+\Delta E_-^{(1)}+\Delta E_-^{(2)}$
where
\begin{eqnarray}
\Delta E_-^{(1)} &=& {1 \over 4}\sum_{\bf q} f_{\bf q}
\frac
{|\delta {\bf b}|^2\left[\varepsilon_0(q)-\hat{h}_{-{\bf q}-{\bf Q}}
\cdot \hat{h}_{-{\bf q}}\varepsilon_0(|{\bf q} - {\bf Q}|)\right]}
{\varepsilon_0^2(|{\bf q} + {\bf Q}|) - \varepsilon_0^2(q)},
\label{DelEminOne} \\
\Delta E_-^{(2)} &=& {1 \over 2}\sum_{\bf q} f_{\bf q}
\frac
{\left[(\delta {\bf b} \cdot \hat{h}_{-{\bf q}-{\bf Q}})
(\delta {\bf b} \cdot \hat{h}_{-\bf q})
\right]
\varepsilon_0(|{\bf q}+{\bf Q}|)}
{\varepsilon_0^2(|{\bf q} + {\bf Q}|) - \varepsilon_0^2(q)},
\label{DelEminTwo}
\end{eqnarray}
and $\Delta E_+^{(i)}$ of the same form as $\Delta E_-^{(i)}$,
but with ${\bf Q} \rightarrow -{\bf Q}$, up to terms that cancel when the
$\Delta E_{\pm}^{(i)}$'s are summed together to form $\Delta E$.

We now proceed to show $\Delta E^{(1,2)}_{\pm}$ are actually independent
of ${\bf Q}$.  Defining $\phi$ as the angle between {\bf Q} and
{\bf q}, and introducing an upper momentum cutoff $\Lambda$,
one finds for large $\Omega$
\begin{eqnarray}
\Delta E_-^{(1)} &=& - \frac{\Omega|\delta {\bf b}|^2}{16\pi^2}
\int_{k_F}^{\Lambda} dq
\frac{q^2}{Q\varepsilon_0(q)}
\int_0^{2\pi}d\phi
\frac{\cos\phi}{2\frac{q}{Q}\cos\phi+1} \nonumber \\
&=& - \frac{\Omega|\delta {\bf b}|^2}{16\pi^2}
\int_{k_F}^{\Lambda} dq
\frac{q^2}{Q\varepsilon_0(q)}
\left(\frac{\pi Q}{q}\right)
\label{Integral1}
\end{eqnarray}
which is manifestly $Q$-independent.  Clearly the same will be true
of $\Delta E_+^{(1)}$.  For the remaining contribution to
$\Delta E$ is is helpful to combine $\Delta E_+^{(2)}$ and
$\Delta E_-^{(2)}$, which can be cast in the form
\begin{eqnarray}
\Delta E_+^{(2)}&+&\Delta E_-^{(2)} =
-\sum_{\bf q} \frac{f_{\bf q}}{\varepsilon_0(q)}
\left\{
\frac
{\delta{\bf b}\cdot({\bf q}+{\bf Q})\delta{\bf b}\cdot {\bf q}
+\delta b_z^2b_z^2}
{\varepsilon_0^2(|{\bf q} + {\bf Q}|) - \varepsilon_0^2(q)}
\right\},
\label{Integral2}
\end{eqnarray}
The term $\delta b_z^2$ in Eq. \ref{Integral2} vanishes upon integration over $\phi$.
For the remaining two terms we write $\delta {\bf b}$ in the form
$$
\delta {\bf b} = \delta b_{\parallel} \hat{Q} +
\delta b_{\perp} \hat{z} \times \hat{Q} +
\delta b_z \hat{z}.
$$
In terms of these quantities, one finds
\begin{eqnarray}
\Delta E_+^{(2)}&+&\Delta E_-^{(2)} =
-\sum_{\bf q} \frac{f_{\bf q}}{\varepsilon_0(q)}
\left\{
\frac
{\left[\delta b_{\parallel}^2-\delta b_{\perp}^2\right]q^2\cos^2\phi
+Qq\delta b_{\parallel}^2\cos\phi}
{\varepsilon_0^2(|{\bf q} + {\bf Q}|) - \varepsilon_0^2(q)}
\right\} \nonumber \\
&=&
-\frac{\Omega}{4\pi^2Q^2}
\int_{k_F}^{\Lambda}dq\frac{q}{\varepsilon_0(q)}
\int_0^{2\pi} \frac
{\left[\delta b_{\parallel}^2-\delta b_{\perp}^2\right]q^2\cos^2\phi
+Qq\delta b_{\parallel}^2\cos\phi}
{2\frac{q}{Q}\cos\phi + 1} \nonumber \\
&=&
-\frac{\Omega}{4\pi^2Q^2}
\int_{k_F}^{\Lambda}dq\frac{q}{\varepsilon_0(q)}
\left\{
q^2\left[\delta b_{\parallel}^2 - \delta b_{\perp}^2 \right]
\left(-\frac{\pi Q^2}{2q^2} \right)
+Qq\delta b_{\parallel}^2\left(\frac{\pi Q}{q} \right)
\right\},
\label{Integral2b}
\end{eqnarray}
which is again manifestly independent of ${\bf Q}$.  We thus see that,
provided $Q < \mu$, the energy required to introduce an oscillation
in the magnetization is independent of the oscillation wavevector.
This indicates that an effective energy functional for the
magnetization should have vanishing coefficient for the quadratic
gradient term -- effectively, a vanishing spin stiffness.  This
contrasts dramatically with the situation we found for $\mu=0$,
where the stiffness diverged as $b_z \rightarrow 0$.

\end{widetext}

\bibliography{DW_HAF8}

\begin{thebibliography}{93}
\bibitem[{not({\natexlab{a}})}]{note_TI}
\bibinfo{note}{In this paper we will use the abbreviation ``TI'' to refer
  specifically to topological insulators protected by time reversal symmetry.}
\bibitem[{not({\natexlab{c}})}]{note1}
\bibinfo{note}{Note that we are referring here to magnetic systems coupled to
  chiral fermions, as opposed to magnets with Dzyalozhinski-Moriya and related
  interactions. Both classes of magnets are characterized as chiral in the
  literature, but to our knowledge they represent distinct systems.}
\end{thebibliography}

\end{document}